\documentclass[acmtog]{acmart}

\setcopyright{acmcopyright}
\acmJournal{TOG}
\acmYear{2018}\acmVolume{37}\acmNumber{4}\acmArticle{104}\acmMonth{8}
\acmDOI{10.1145/3197517.3201295}


\graphicspath{{figures/images/}} 

\usepackage{color}
\newcommand{\ca}[1]{{\color{black}            {#1}}}

\usepackage{color}
\definecolor{green}{rgb}{0, 0.5, 0}
\definecolor{orange}{rgb}{0.8, 0.6, 0.2}
\definecolor{red}{rgb}{1.0, 0.0, 0.0}
\definecolor{teal}{rgb}{0.0, 0.4, 0.4}
\definecolor{purple}{rgb}{0.65,0,0.65}
\definecolor{saffron}{rgb}{0.95,0.75,0.2}
\definecolor{turquoise}{rgb}{0.0,0.5,0.5}

\usepackage[ruled,vlined,linesnumbered]{algorithm2e}

\usepackage{multirow}

\usepackage[normalem]{ulem}

\usepackage{mathtools}
\usepackage{amsmath, amsthm, amssymb, amsfonts, amsopn}
\usepackage{graphicx} 
\usepackage{textcomp}
\usepackage{xfrac}
\usepackage{bbm}

\usepackage{algorithmic}

\usepackage{overpic}
\usepackage{subfig}
\usepackage{wrapfig}

\newcommand{\hidecomment}[1]{}

\newcommand{\pcite}{\protect\cite}

\begin{document}

\title{Object-Aware Guidance for Autonomous Scene Reconstruction}

\author{Ligang Liu}
\affiliation{%
	\institution{University of Science and Technology of China}
}
\author{Xi Xia}
\affiliation{%
	\institution{University of Science and Technology of China}
}
\author{Han Sun}
\affiliation{%
	\institution{University of Science and Technology of China}
}
\author{Qi Shen}
\affiliation{%
	\institution{University of Science and Technology of China}
}
\author{Juzhan Xu}
\affiliation{%
	\institution{Shenzhen University}
}
\author{Bin Chen}
\affiliation{%
	\institution{Shenzhen University}
}
\author{Hui Huang}
\affiliation{%
	\institution{Shenzhen University}
}
\author{Kai Xu}
\authornote{Kai Xu is the corresponding author (kevin.kai.xu@gmail.com)}
\affiliation{%
	\institution{Shenzhen University}
}
\affiliation{%
	\institution{National University of Defense Technology}
}

\renewcommand\shortauthors{L. Liu, X. Xia, H. Sun, Q. Shen, J. Xu, B. Chen, H. Huang and K. Xu}

\begin{abstract}
\ca{
To carry out autonomous 3D scanning and online reconstruction of unknown indoor scenes, one has to find a balance between global exploration of the entire scene and local scanning of the objects within it.
In this work, we propose a novel approach, which provides object-aware guidance for autoscanning, for exploring, reconstructing, and understanding an unknown scene within \emph{one navigation pass}.
%
Our approach interleaves between object analysis to identify the \emph{next best object} (NBO) for global exploration, and object-aware information gain analysis to plan the \emph{next best view} (NBV) for local scanning.
First, an objectness-based segmentation method is introduced to extract semantic objects from the current scene surface via a multi-class graph cuts minimization.
%
Then, an object of interest (OOI) is identified as the NBO which the robot aims to visit and scan.
%
The robot then conducts fine scanning on the OOI with views determined by the NBV strategy.
When the OOI is recognized as a full object, it can be replaced by its most similar 3D model in a shape database.
The algorithm iterates until all of the objects are recognized and reconstructed in the scene.
Various experiments and comparisons have shown the feasibility of our proposed approach.
}

\end{abstract}

\citestyle{acmauthoryear}
\setcitestyle{square}



\begin{teaserfigure}
   \begin{overpic}[width=1.0\textwidth,tics=100]{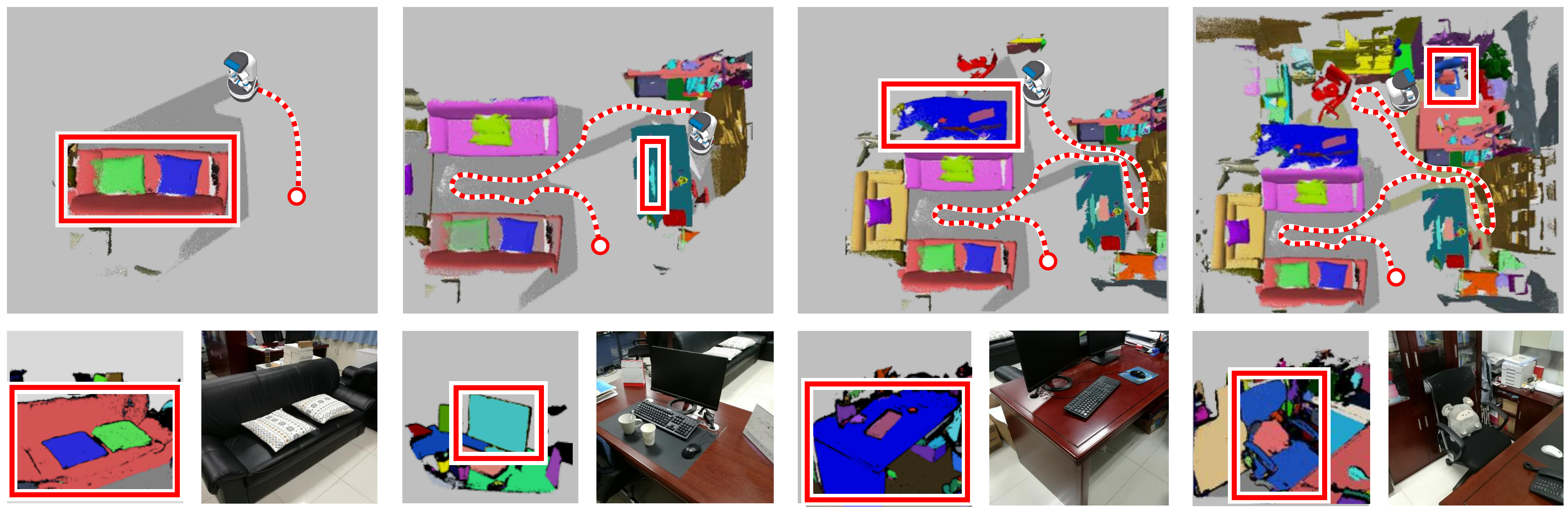}
    \put(11.5,-1){\small (a)}
    \put(36.7,-1){\small (b)}
    \put(61.8,-1){\small (c)}
    \put(87.2,-1){\small (d)}
   \end{overpic}
   \caption{
Autonomous scene scanning and reconstruction on a real office scene using our object-aware guidance approach. In each column (a)-(d), the object marked with the \ca{red} rectangular frame is the \ca{object-of-interest (OOI)}. The upper row shows the navigation path (in dotted red) with previous scanning views (shown as \ca{white} dots) and the current position of the robot. The objects in different colors are the reconstructed objects in the scene.
The bottom row shows the depth data (left) and the RGB image (right) from the current view of the robot.
%
Our approach achieves both global path planning and local view planning on-the-fly within one single navigation pass and obtains the reconstructed scene with semantic objects (d).
   }
   \label{fig:teaser}
\end{teaserfigure}
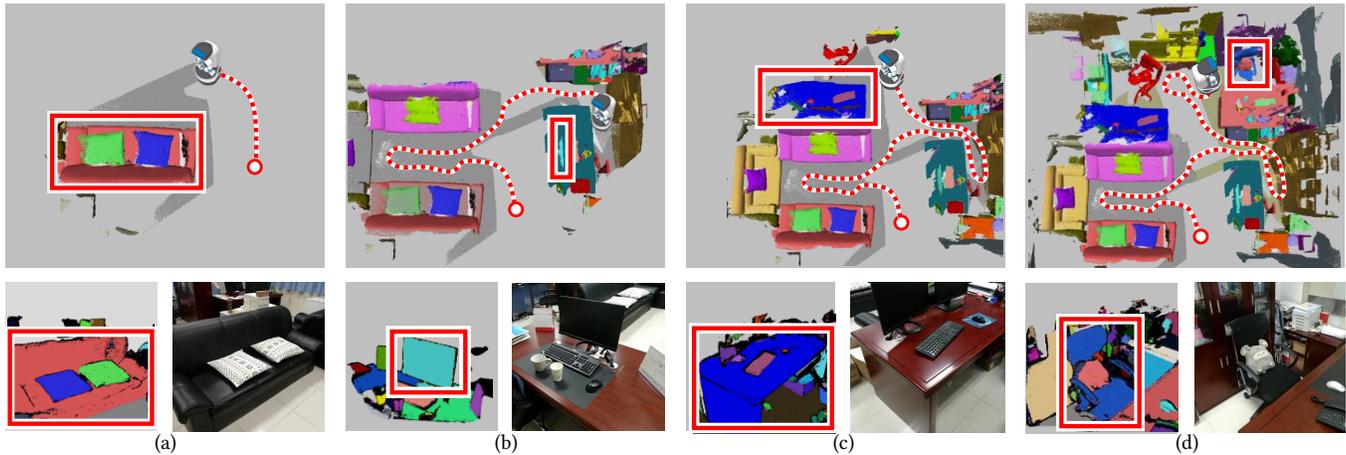

\maketitle

\section{Introduction}
\label{sec:intro}

\ca{
%
In recent years autonomous 3D scanning and reconstruction of indoor scenes by mobile robots with RGB-D depth sensors have become crucial research areas in both the robotics and graphics communities~\cite{Xu15,Ramanagopal2016,Xu2016}.
}

To automatically explore and reconstruct unknown scenes, a robotic system demands two integrated levels of autonomous navigation planning.
On the one hand, \emph{path planning} aims to expand knowledge of the scene by creating a map of \emph{global} spatial information about it and localizing itself within the map before planning a next-best \emph{robot position}~\cite{Fan2016}.
On the other hand, \emph{view planning}, also known as next-best-view (NBV) planning~\cite{Wu2014}, aims to describe a \emph{sensor viewpoint }, which provides the best sensory input to create high-quality 3D models by fusing \emph{local} geometric information from visible surfaces.
%

However, existing autoscanning systems usually regard global exploration and local scanning as separate problems.
\ca{
%
They generally acquire scene maps and 3D scene data in the first navigation and then either perform offline analysis and reconstruction on the acquired data or conduct detailed scanning of individual objects in the second pass.
%
Automatically achieving both tasks in one navigation pass remains challenging for autonomous scene reconstruction.
}

\begin{figure}[t!] \centering
	\begin{overpic}[width=1.0\linewidth,tics=10]{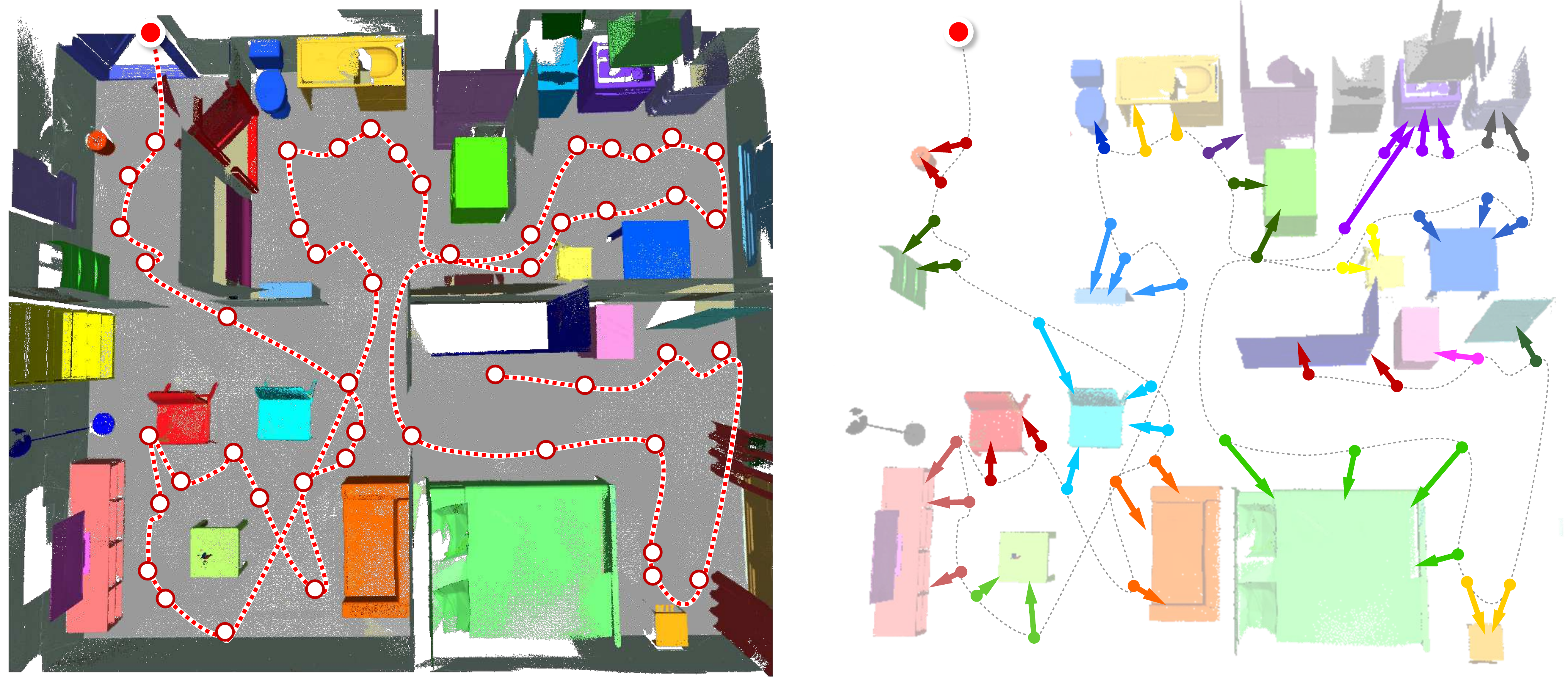}
	\end{overpic}
    \caption{
    The planned global path (left) and local views (right) on autonomous scanning an unknown scene produced by our approach. The red point near the top left is the starting point where the robot enters the scene. (Left) the dotted red line shows the navigation path of the robot in the scene; the white points on the path are the positions where the robot goes to for local scanning;
    (Right) each arrow shows the view direction of the robot at its position, pointing to the corresponding object it is scanning. The arrows which correspond to the same object are shown in same color.
    }
    \label{fig:pathnbv}
    \vspace{-7pt}
\end{figure}

\ca{In this paper, we propose a novel autoscanning approach, which provides \emph{object-aware guidance} for allowing robots to simultaneously complete both global path planning and local detailed view planning on-the-fly, within \emph{one} single navigation pass.
The robot identifies semantic objects, visits them one by one, and simultaneously carries out active scans of the currently visiting object (Fig.~\ref{fig:teaser}).
This is mainly inspired by the observation that when a human scans an indoor scene with a hand-hold scanner, she is inclined to identify one object and then scan it completely before she goes to scan the next one.
}
%
Likewise, while exploring the scenes in our system, the robot first identifies the object with the largest degree of recognition from its current point of observation, which is called the \emph{next-best-object} (NBO), and sets it as the \emph{object-of-interest} (OOI). Then, the robot visits and scans the OOI with NBVs that are driven by increasing its recognition degree. After the robot completes the reconstruction of the current OOI, it goes on identifying and scanning the next OOI. This sequential visiting and scanning of all of the objects constitutes one navigation path of the robot in the scene.
%
%

The core of our approach is an iterative \emph{identification-and-planning} algorithm, which interleaves between object-level shape analysis to identify the NBOs for global navigation, and object-aware information gain analysis to plan the NBVs for local scanning.
First, we present a notion of model-driven \emph{objectness}, which is based on prior knowledge of 3D shapes provided by a 3D model database. 
%
\ca{
Second, we develop an objectness based segmentation method to extract objects via a multi-class graph cuts optimization.
%
%
The robot then visits the identified OOI and scans it with views that are determined by the NBV strategy, which is based on increasing the information gain in order to recognize the OOI.
When the OOI is recognized as a full object (with a very high objectness score), the most similar 3D model is retrieved from the database and inserted into the scene to replace it.
Our algorithm interleaves NBO and NBV estimations until all objects are reconstructed in the scene (Fig.~\ref{fig:pathnbv}).
%
%
}

%

The contributions of our work include:
\begin{itemize}
  \item a unified robotic autoscanning system, which enables on-the-fly exploration, reconstruction, and understanding of unknown scenes in one navigation pass;
  \item a model-driven objectness metric for measuring the similarity and completeness of segmented components from objects in the 3D model database;
  \item an objectness-based segmentation method via a multi-class graph cuts minimization, which couples segmentation and recognition in the same optimization;
  \item an objectness-based exploration and scanning strategy based on the NBO and NBV schemes, which accomplishes an object-guided autonomous scene reconstruction.
\end{itemize}

As far as we know, our approach is the first to process a stream of scanned depth frames on-the-fly in order to perform autonomous exploration and semantic reconstruction of unknown scenes in one single navigation pass.
%
\ca{
Our work is quite different from the classic treasure hunting problem in robotics, because we aim to simultaneously scan the scene and recognize various objects in it through online reconstruction and analysis.}
Our algorithm is integrated with an autonomous robotic system for scene scanning, which is operated by a personal robot holding a depth camera.
A large number of experiments and comparisons have been conducted to evaluate the feasibility and effectiveness of our proposed approach.
%

\begin{figure*}[t!] \centering
	\begin{overpic}[width=1.0\linewidth,tics=10]{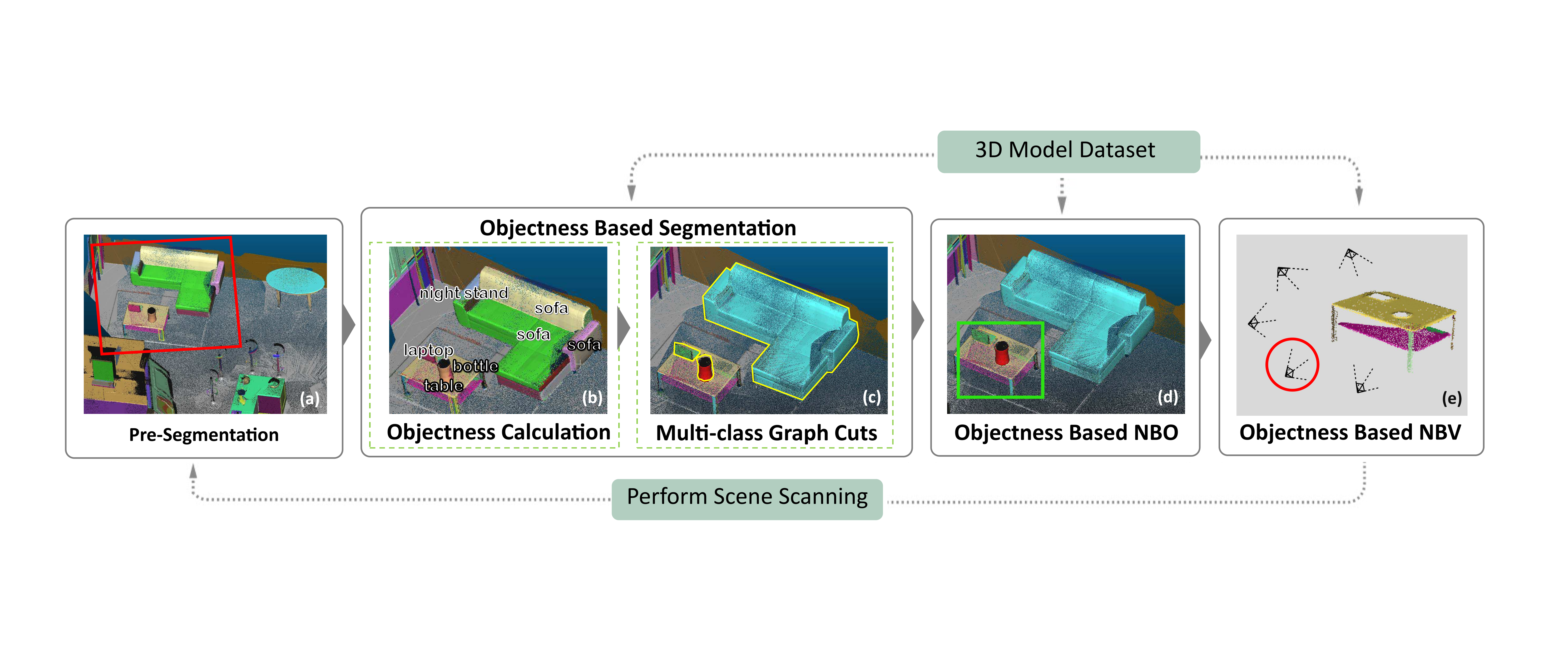}
	\end{overpic}
    \caption{Pipeline of our object-guided autonomous scene scanning and reconstruction approach.}
    \label{fig:overview}
    \vspace{-7pt}
\end{figure*}

\section{Related work}
\label{sec:related}

\paragraph{Autonomous scene exploration and scanning}
With the emergence and rapid development of personal robotics and commodity depth cameras, there have been a large variety of works on exploring large-scale indoor scenes and scanning single objects via autonomous scanning systems~\cite{Krainin2012Autonomous,Wu2014,Charrow2015,Xu2017}.
The autoscanning system is generally equipped with a mobile robot with a fixed camera or an articulated robotic arm holding a depth camera~\cite{Krainin2012Autonomous,Charrow2015,Kriegel2012}.
For large-scale indoor scenes, the robot is expected to build the map of the scene and localize itself in the map while exploring the scene based on the Simultaneous Localization and Mapping (SLAM) techniques~\cite{Engelhard2011,Salas-Moreno2013,Whelan2015}.
For individual objects, the robot is expected to fine scan the objects with fine grained geometric details and then reconstruct their geometries with a careful plan of scanning views~\cite{Kriegel2012,Wu2014,Xu2016}.
However, the systems for autonomous scanning the scenes and the systems for autonomous scanning the objects are generally separated. There is little work on coupling scans of both scenes and objects in one navigation pass.
\ca{The work in~\cite{Xu2017} allows one-pass navigation for autoscanning, however, object recognition is missing in their work.}
We develop a novel autoscanning approach which can accomplish both the exploration of an unknown scene and the semantic reconstruction of the objects in it, via processing on-the-fly acquired depth data.

\paragraph{Scene analysis and understanding}
Scene reconstruction requires the high-level analysis and understanding on the objects and their mutual spatial relationships in the scene, to reveal the composition and structure of the scene~\cite{Fisher2012,Salas-Moreno2013,Valentin2015}.
Data-driven methods, which utilize a 3D model database with object level segmentation and  semantic labels to assist object extraction, understanding, and analysis, have attracted more attention in recent years and enabled extracting structural and contextual relationship between objects~\cite{Nan2012,Salas-Moreno2013,Zhang2015Online,Xu2016}.
Our work also takes a data-driven approach. Specifically, we retrieve the similar models in the database as shape priors for identifying and recognizing objects in the scene and use them as the guidance for robot's movements on both global path planning and local scan view planning.
\ca{Our method is different from~\cite{Xu15} since it relies on robot interaction for object extraction, while ours achieves that by object recognition.}

\paragraph{Global path planning}
In order to capture the global structure of the scene, robot path planning has to be considered to work with robust camera localizatoin and scan registration with SLAM, which can be achieved by jointly minimizing the the uncertainty of both scene mapping and camera localization~\cite{Thrun2002Robotic}.
There have been a bunch of works on this problem. Most methods require the data to be overlapped sufficiently for the ease of frame-to-frame registration. Some methods focus on handling the drift issues and/or the loop closure problem due to the registration errors~\cite{Agarwal2010Bundle,Choi2015,Newcombe2012KinectFusion,Ramanagopal2016,Zeng2016}.
%
The recent work of~\cite{Xu2017} presents a time-varying tensor field based scheme to simultaneously compute smooth movement for both robot path and camera trajectory.
Nevertheless, this scheme can only perform pass-through scanning along with the robot movement rather than detailed scanning on individual objects and no object recognition is performed.
Different from existing works, our approach identifies the objects and uses them as a high level guidance to explore the scene, which is inspired from the object-aware attention mechanism that humans focus attention selectively on identified objects being observed.

\paragraph{Local view planning}
The selection of view directions (NBVs) for a robot sensor is critical for autonomous scanning for capturing geometry of 3D objects.
The goal is to reduce the recognition uncertainty and recover the surface of an object with a minimal number of scanning views.
Many NBV algorithms have been developed for actively acquiring and scanning 3D objects~\cite{Wu2014,Xu2016} as well as 3D scenes~\cite{Low2006,Xu15,Fan2016}.
Different from existing methods, we take advantage of the retrieved 3D shapes in the database as shape prior and perform an object-aware information gain analysis for planning the NBVs for local scanning. The key in our approach is that we employ the object identification and view planning in a couple manner.

\IncMargin{0.5em}
\begin{algorithm} [t]
\caption{Object-guided auto scene reconstruction}
\label{algo:timestep}
\SetCommentSty{textsf}
\SetKwInOut{AlgoInput}{Input}
\SetKwInOut{AlgoOutput}{Output}

\SetKwFunction{CurrentFrame}{AcquiredDepthData}
\SetKwFunction{PreSeg}{PreSegmentation}
\SetKwFunction{PostSeg}{PostSegmentation}
\SetKwFunction{NBO}{NextBestObject}
\SetKwFunction{NBV}{NextBestView}

\AlgoInput{ Initial viewpoint of robot $V$ }
\AlgoOutput{ Reconstructed scene $\mathcal{S}$ }
Initialization:  $\mathcal{S} \leftarrow \varnothing$,  $\mathcal{T} \leftarrow \CurrentFrame{V}$\;
\Repeat {$\mathcal{T} = \varnothing$} {
    $\mathcal{C} \leftarrow$ \PreSeg{$\mathcal{T}$}\;
    $\mathcal{R} \leftarrow$ \PostSeg{$\mathcal{C}$}\;
    $\gamma \leftarrow$ \NBO{$\mathcal{R}$}\;
    $V \leftarrow$ \NBV{$\gamma$}\;
    $\mathcal{T} \leftarrow \mathcal{T} \cup \CurrentFrame{V}$\;
    \If{$\gamma$ is a complete object} {$\mathcal{S} \leftarrow \mathcal{S}\cup \{ \gamma \}$, $\mathcal{T} \leftarrow \mathcal{T} \setminus \{ \gamma \}$}
}
\end{algorithm}
\DecMargin{0.5em}

\section{Overview}
\label{sec:overview}

Fig.~\ref{fig:overview} shows the pipeline of our object-guided autoscanning system.
%
\ca{The movement of the robot $\Omega$ is described as a sequence of steps on its positions or its sensor viewpoints.
After $\Omega$ enters an unknown indoor room, it starts to acquire a stream of raw RGB-D data as the input.
An underlying RGB-D SLAM framework runs to fuse the acquired depth data with the current scene surface $\mathcal{T}$.
%
We denote $\mathcal{S}$ ($\mathcal{S}=\varnothing$ in the beginning) as the reconstructed scene, which includes the previously scanned objects  from prior steps.
%
%
The database of 3D indoor models is denoted as $\mathcal{M}$ .

At each step, $\mathcal{T}$ is first segmented into a set of pre-segmented near-convex components (Fig.~\ref{fig:overview}(a)).
Based on an objectness measurement (Fig.~\ref{fig:overview}(b)), the pre-segmented components are merged into a set of post-segmented objects (Fig.~\ref{fig:overview}(c)) via a multi-class graph cuts minimization, which are adopted as high-level guidance to plan the movement of $\Omega$.
%
%

%
The most salient object is chosen as the OOI $\gamma$ (Fig.~\ref{fig:overview}(d)) and then $\Omega$ moves its position to $\gamma$ and starts actively scanning it.
With the assistance of similar models of $\gamma$ in $\mathcal{M}$, the optimal NBV $V$ is chosen for the next observation and scan (Fig.~\ref{fig:overview}(e)). 
When $\gamma$ is recognized as a complete object (with a very high objectness score), it is then replaced by its most similar 3D model in $\mathcal{M}$. 
}

This repeats until all of the post-segmented objects in $\mathcal{T}$ have been processed, resulting in a full reconstruction $\mathcal{S}$ of the scene. Algorithm~\ref{algo:timestep} summarizes the whole autoscanning process.

\section{Objectness-based segmentation}
\label{sec:segmentation}

%

\subsection{Pre-segmentation}
\label{sec:pre-segmentation}

\paragraph{Underlying SLAM framework}
Underlying our system we run a GPU version of the dense RGB-D SLAM framework~\cite{Whelan2015} to record robot's trajectories and camera transformations.
This SLAM framework is based on a surfel-based fusion method with a global loop closure optimization, which is capable of capturing comprehensive dense globally consistent surfel-based maps of room scale environments with high accuracy.
The current scene surface $\mathcal{T}$ is updated by fusing the current depth data at each time step.
%



\paragraph{Pre-segmentation}
\ca{
An incremental segmentation algorithm~\cite{tateno2015} is run on top of the SLAM framework, which separately segments the acquired depth data and then incrementally merges the obtained segments within a unified global segmentation map by means of the estimated camera pose.}
The method is able to segment $\mathcal{T}$ into near-convex components (like legs and arms of chairs), which are called \emph{pre-segmented components} and denoted as $\mathcal{C} = \{ c_i, i=1,2,\cdots, n_c \}$, in realtime (Fig.~\ref{fig:overview}(a)).
Note that some pre-segmented components may not exist as complete semantic components from the models if they are just partially scanned or if the acquired data is incomplete due to occlusion.



\begin{figure}[t!] \centering
	\begin{overpic}[width=1.0\linewidth,tics=10]{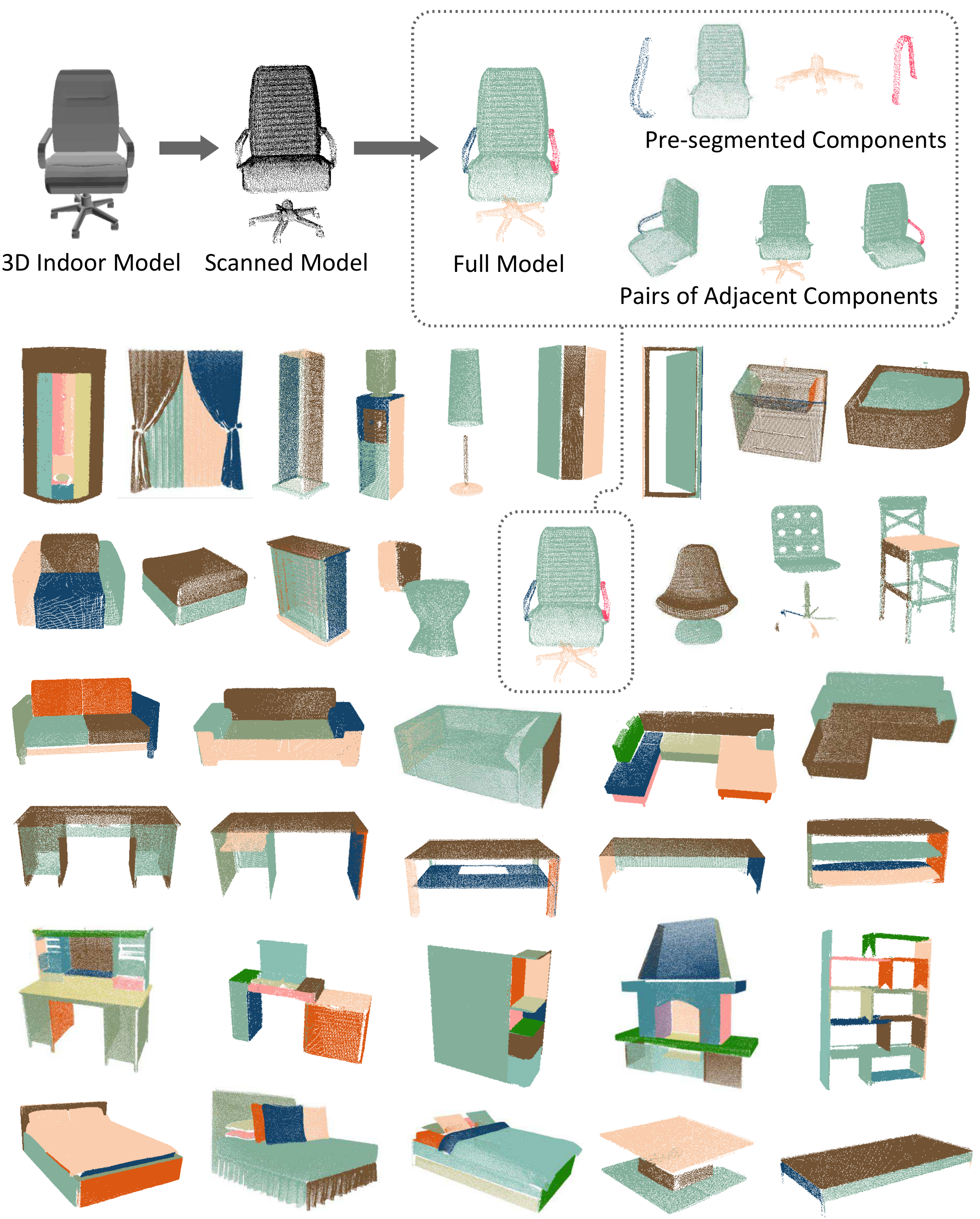}
	\end{overpic}
    \caption{The database $\mathcal{M}$ of 3D indoor models.
    For each model, such as the chair, we virtually scan it and put it into $\mathcal{M}$. Then, we segment it into pre-segmented components and put them into $\mathcal{M}$.
    We also put the pairs of adjacent pre-segmented components into $\mathcal{M}$.
    }
    \label{fig:database}
    \vspace{-7pt}
\end{figure}

\subsection{Model-driven objectness}
\label{sec:objectness}

\paragraph{3D model database}
We construct a 3D model database $\mathcal{M}$ (Fig.~\ref{fig:database}) to provide $\Omega$ with prior knowledge of 3D shapes and endow it with the ability to recognize and identify semantic objects in $\mathcal{T}$.
We first collect $n_l$ classifications of 3D indoor models $\mathcal{M}^*$ , such as chairs, tables, sofas, beds, and book shelves, etc. Each classification is assigned with a label $L \in \{1,2,\cdots,n_l\}$.
As the 3D data captured by the depth sensors is of low quality with noises, we convert these clean models into point data to support more accurate and precise online partial matching and object recognition.
Specifically, we virtually scan each model $Z \in \mathcal{M}^*$, with a label, denoted as $L(Z)$, into 3D point data $m$.
Then, we segment $m$ into pre-segmented components using the algorithm~\cite{tateno2015}.
Moreover, for any pair of adjacent pre-segmented components of $m$, we merge them into one larger component.
We designate $m$, the pre-segmented components, and the pairs of adjacent components with the same label $L(Z)$ and put all of them into $\mathcal{M}$.

%
\ca{
It is worthwhile pointing out that we put all components as well as \emph{pairs of adjacent components} into $\mathcal{M}$.
The insight is that the merging of two adjacent components allow our proposed graph cuts optimization (Section~\ref{sec:graphcuts}) to merge multiple adjacent pre-segmented components in $\mathcal{C}$ into more complete objects, hence significantly enhancing the identification and recognition of the objects.
}

\paragraph{Similar model set}
\ca{
For each $c \in \mathcal{C}$, we search for a few models in $\mathcal{M}$ that are most similar to $c$ and use them as candidates for recognition.
}
As $c$ might be incomplete due to occlusion, we adopt a partial matching method to find the similar models.
First, we uniformly sample $n_p$ keypoints ($n_p =500$) from each $m \in \mathcal{M}$ and $c$ respectively using the Lloyd's algorithm~\cite{Lloyd1982Least}.
Second, we utilize a learning-based 3D shape descriptor, 3DMatch~\cite{Zeng2016}, which is learnt from a real scanned depth data of indoor scenes. 
%
Then, we assign each keypoint with the three closest cluster centers according to the codebook of $\mathcal{M}$ and compute histograms of every cluster center. In addition, we use a spatially-sensitive bag of words (BoW) to overcome the drawback by considering the spatial position among the keypoints.
Thus, we can quickly compute a similar model set $M(c) = \{m_1, m_2, \cdots, m_{n_s} \} \subset \mathcal{M}$ with $n_s$ models from $\mathcal{M}$ that are most similar to $c$ ($n_s = 5$ in our implementation).

\begin{figure}[t!] \centering
	\begin{overpic}[width=1.0\linewidth,tics=10]{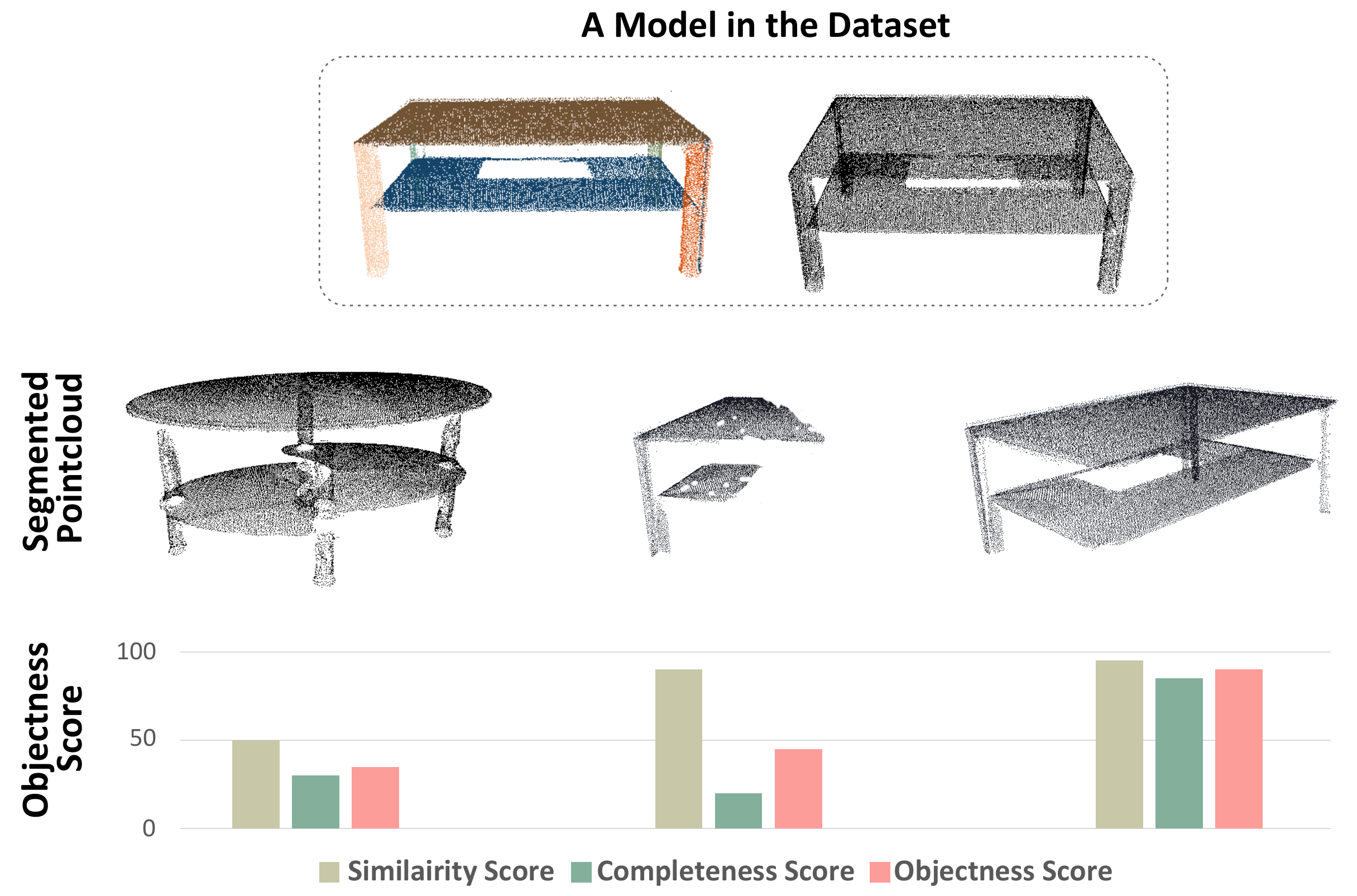}
	\end{overpic}
    \caption{The objectness score measures how much a segmented pointcloud in the current scene surface $\mathcal{T}$ matches a model in the database $\mathcal{M}$ in both similarity and completeness.
    }
    \label{fig:objectness-score}
    \vspace{-7pt}
\end{figure}

\paragraph{Objectness}
For two 3D data points $X$ and $Y$, we define the matching rate of $X$ according to $Y$ as:
$$ d(X,Y) = \frac{1}{n_p} \sum_{i=1}^{n_p} {d(x_i, Y)} $$
where
$$d(x_i, Y) = \min \limits_{j=1,\cdots,n_p} \lVert{x_i-y_j}\rVert^2, $$
$\{x_i\}_{i=1}^{n_p}$ and $\{y_i\}_{i=1}^{n_p}$ are sampled keypoints of $X$ and $Y$, respectively.

For $c$ and $m \in M(c)$, $d(c,m)$ measures the \emph{similarity} of $c$ according to $m$, and $d(m,c)$ measures the \emph{completeness} of $c$ according to $m$.
An objectness score is defined to measure how much $c$ matches $m$ in both similarity and completeness as (Fig.~\ref{fig:objectness-score}):
\begin{equation}\label{equ:objectness}
  O(c,m) = \exp\left[ { -\frac{1}{ Diag(c)} ( d(c,m) + d(m,c) )^{\frac{1}{2}} }\right]
\end{equation}
where $Diag(c)$ is the diagonal length of the bounding box of $c$.
%

\subsection{Post-segmentation: objectness-based segmentation}
\label{sec:graphcuts}

\begin{figure}[t!] \centering
	\begin{overpic}[width=1.0\linewidth,tics=10]{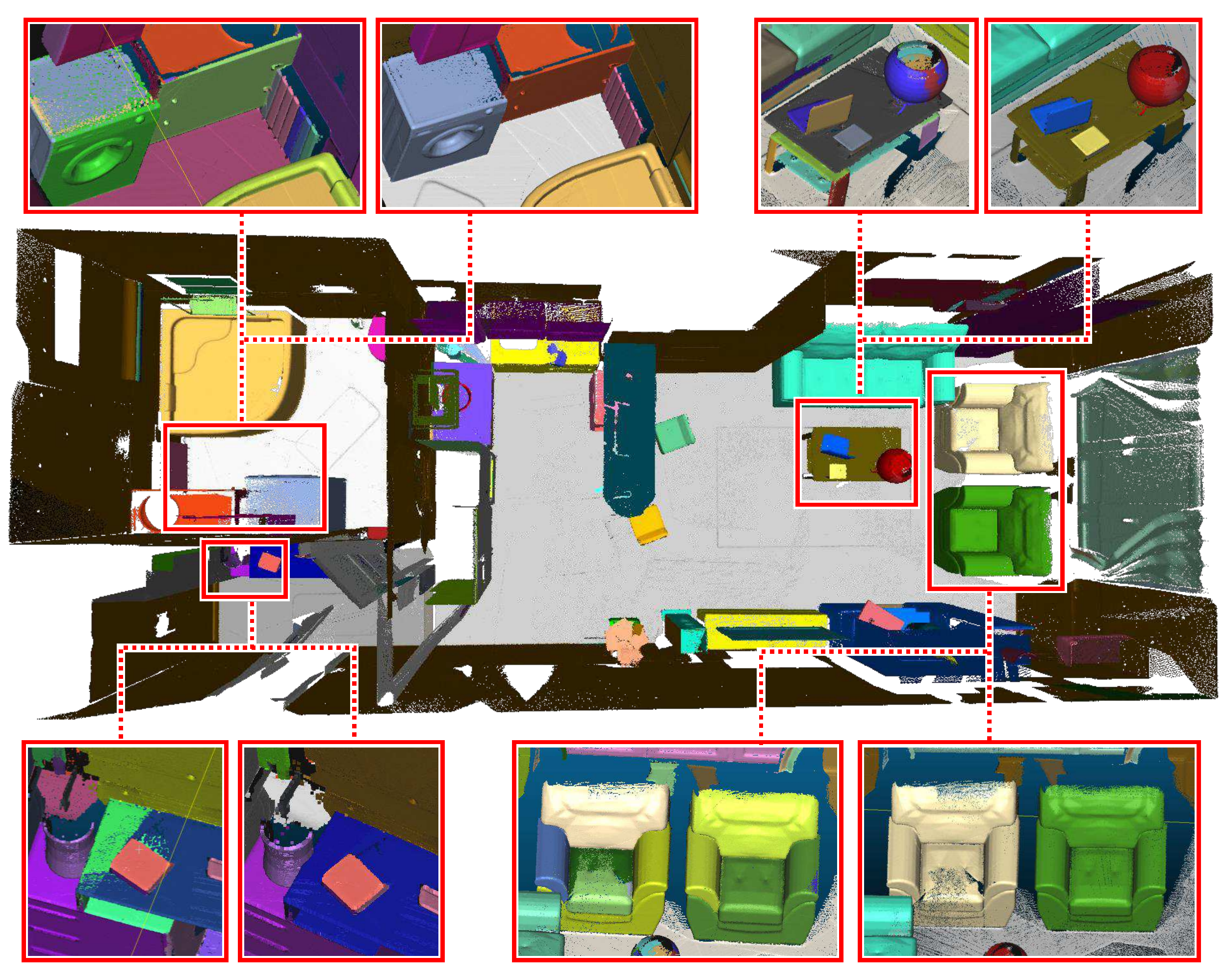}
	\end{overpic}
    \caption{
    Four regions in the scanned scene (middle row) are zoomed and their segmentation results of pre-segmentation (left) and post-segmentation (right) are shown side by side for comparisons.
    }
    \label{fig:graph-cuts-segmentation}
    \vspace{-7pt}
\end{figure}

%
We further develop a post-segmentation technique to refine the pre-segmentations and obtain object-level segmentations (Fig.~\ref{fig:graph-cuts-segmentation}). This is an objectness-based segmentation that is carried out by integrating the objectness measurement with the recognition rate in a multi-class graph cuts optimization, which efficiently \emph{couples} both segmentation and recognition in the same optimization.

\paragraph{Formulation}
The goal is to assign each pre-segmented component $c \in \mathcal{C}$ with a label $l_c \in \{1,2,\cdots,n_l\}$ so that adjacent components from the same object have the same labels, thus allowing them to  be merged into a more complete object.
We first build an adjacency graph for all components in $\mathcal{C}$. This is denoted as $\mathcal{G}_c = (\mathcal{V}_c, \mathcal{E}_c)$ with the nodes representing the pre-segmented components and the edges indicating the component adjacency relations.
Based on the component graph, we compute the post-segmentation using the following graph cuts minimization:
\begin{equation}\label{equ:graph-cuts}
  \min \limits_{L=\{ l_c \}} E(L) = \sum \limits_{c \in \mathcal{V}_c} E_D (l_c)
     + \sum \limits_{(c,d) \in \mathcal{E}_c} E_S(l_c, l_d)
\end{equation}
where $l_c$ and $l_d$ are the labels for $c$ and $d$, respectively, and $E_D ( l_c )$ and $E_S ( l_c, l_d )$ are the data term and smoothness term, respectively.


\paragraph{Data term}
The data term penalizes the probability of $c$ not being labelled as $l_c$ as:
\begin{equation}\label{equ:data-term}
  E_D ( l_c ) = \min \limits_{m \in M(c), l_c = L(m) } (1-O(c,m)).
\end{equation}

\paragraph{Smoothness term}
The smoothness term is defined as the probability of two adjacent components $c$ and $d$ belonging to the same object, i.e., labeled differently as:
\begin{equation}\label{equ:smooth-term}
  E_S ( l_c, l_d ) =
  \begin{cases}
      \max \limits_{ m \in M(c \cup d) } O(c \cup d,m), &\text{if } l_c \neq l_d,\\
      0, &\text{if } l_c = l_d.
  \end{cases}
\end{equation}
%


As our multi-class graph cuts technique couples segmentation and recognition in the same optimization, the post-segmentation method can effectively merge adjacent pre-segmented components into more complete objects, as shown in Fig.~\ref{fig:graph-cuts-segmentation}.

\ca{
Our multi-class graph cuts optimization tends to merge adjacent pre-segmented components based on the objectness metric. As we also put pairs of adjacent components into the database, our optimization may merge multiple adjacent pre-segmented components into more complete objects. Thus our method can efficiently merge those adjacent pre-segmented components by significantly enhancing identification and recognition of the objects. This can be hardly achieved by existing methods.
}

\section{Objectness-based reconstruction}
\label{sec:reconstruction}

Using the objectness-based segmentation technique, we obtain a set of post-segmented objects, denoted as $\mathcal{R} = \{ r_1, r_2, \cdots, r_{n_r} \}$, in the current scene surface $\mathcal{T}$.

\begin{figure}[t!] \centering
	\begin{overpic}[width=1.0\linewidth,tics=10]{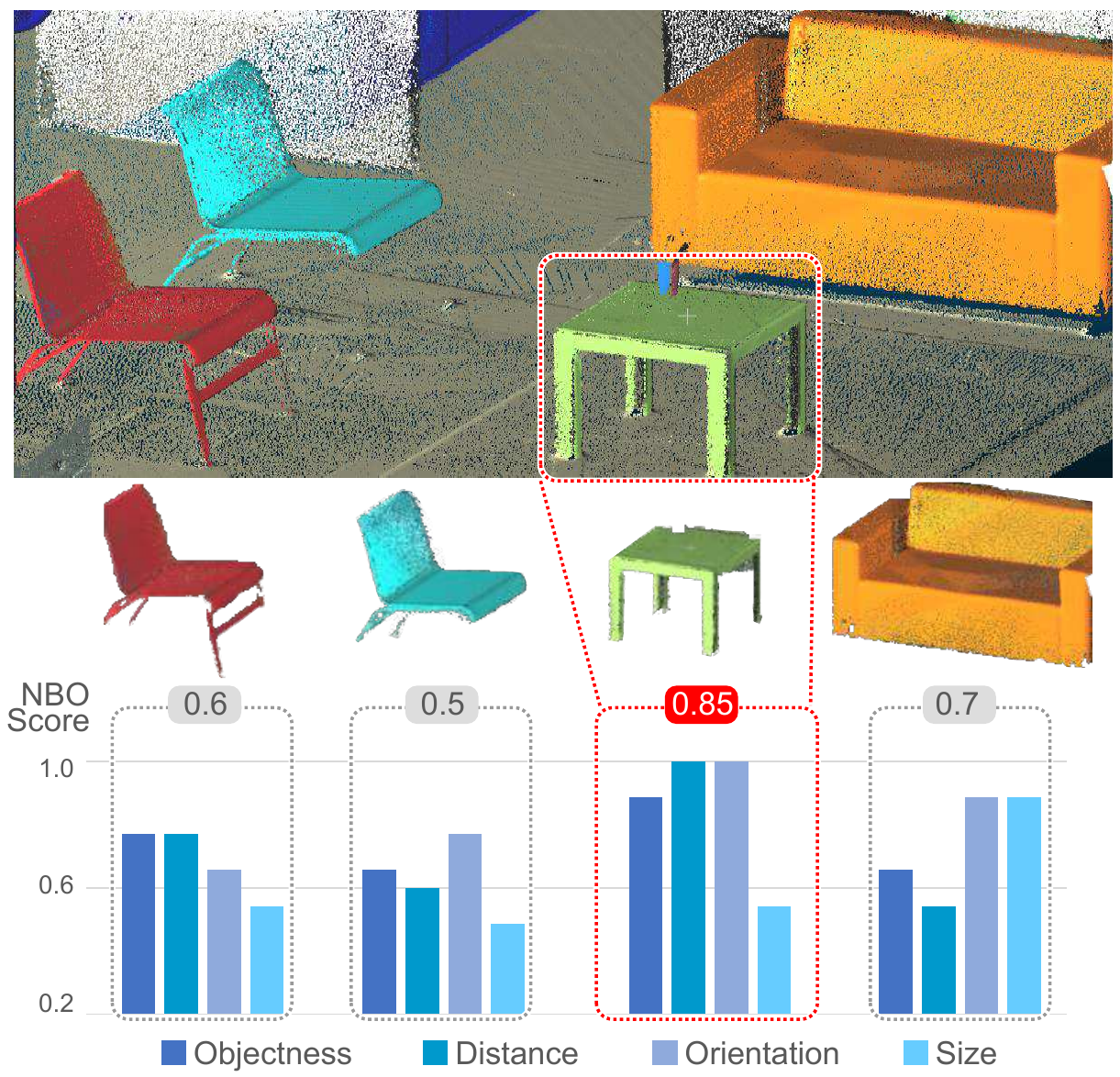}
	\end{overpic}
    \caption{
    The NBO scores of four post-segmented objects are shown. The table marked by the red rectangular frame has the highest NBO score and is selected as the OOI.
    }
    \label{fig:NBO}
    \vspace{-7pt}
\end{figure}

\subsection{The next best object (NBO)}


The robot $\Omega$ needs to identify the OOI, denoted as $\gamma$, in $\mathcal{R}$ as its next object to visit.
When the scores for objectness and visual saliency are added together, the post-segmented object with the largest sum is chosen as $\gamma$ in $\mathcal{R}$ (Fig.~\ref{fig:NBO})
\begin{equation}\label{equ:NBO}
\gamma = \arg \max_{r \in \mathcal{R}}  O(r) + S(r)
\end{equation}
where \ca{$O(r)=\max_{m \in M(r)}  O(r,m)$ measures how much $r$ could be an object in $\mathcal{M}$ }
and $S(r)$ is the saliency score of $r$ according to the robot's current view direction $V$.
The saliency score contains three parts:
$$
S(r) = w_z S_z (r) + w_e S_e (r) +  w_d S_d (r)
$$
where the distance score
$S_z (r) = \exp \left[{-(C(r)-P_{\Omega})/{\max \limits_{\bar{r} \in \mathcal{R}}(C(\bar{r})-P_{\Omega})}} \right] $
measures the distance from the center $C(r)$ of $r$ to the position $P_{\Omega}$ of the robot $\Omega$.
The orientation score $S_e (r) = (C(r)-P_{\Omega}) \cdot V $ measures the angle between the orientation of $c$ and the view direction of $\Omega$.
The size score $S_d (r) =  Area(r)/{\max \limits_{\bar{r} \in \mathcal{R}} Area(\bar{r})} $, \ca{where $Area(r)$ is the area of $r$,} measures the size of $r$, and $w_z, w_e, w_d$ are weights ($w_z=1.5, w_e=1, w_d=1$ by default).


\begin{figure}[t!] \centering
	\begin{overpic}[width=1.0\linewidth,tics=10]{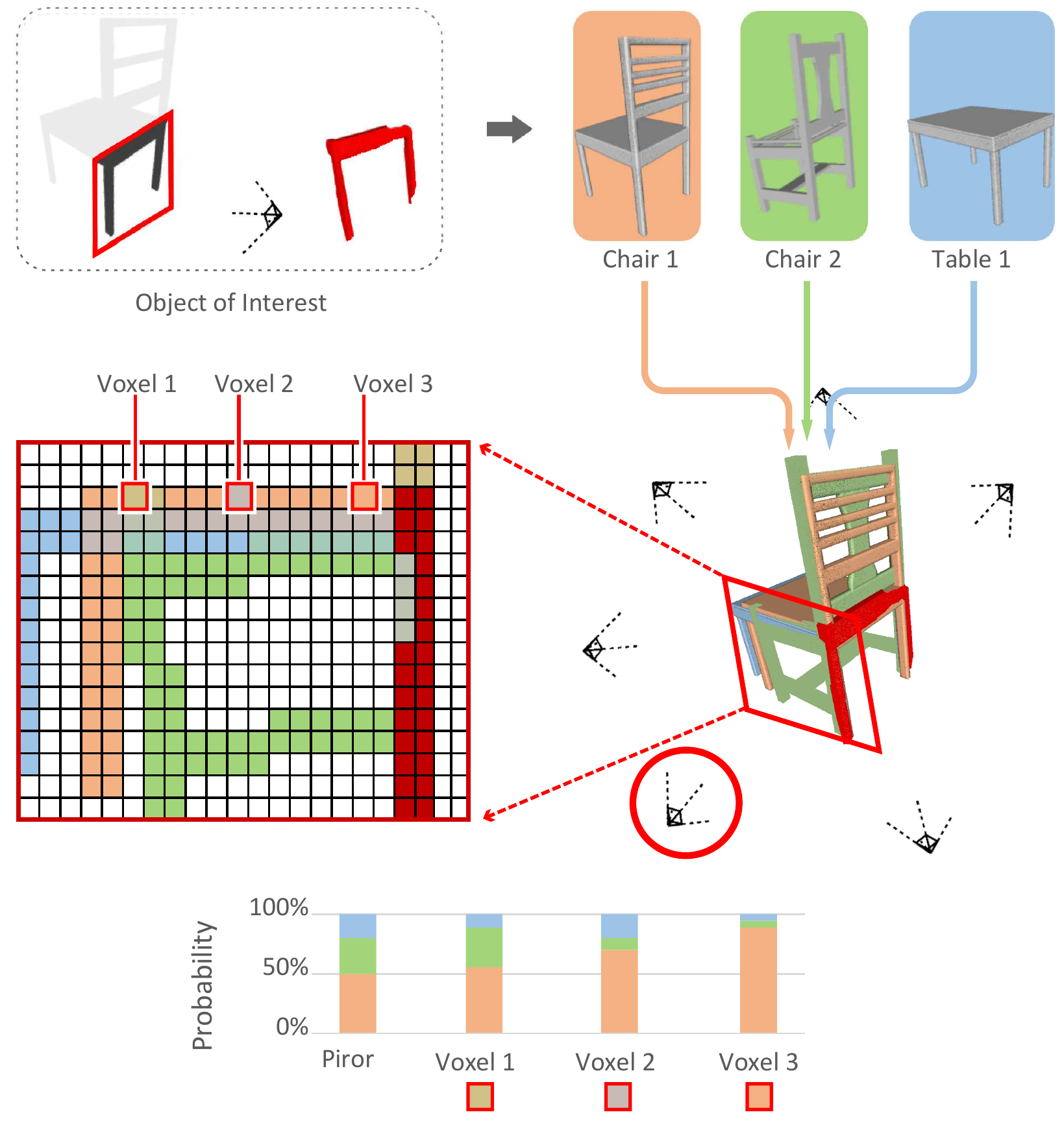}
	\end{overpic}
    \caption{
    An illustration of the computation of NBV.
    For the current OOI $\gamma$ (top-left), 3 model candidates Chair 1, Chair 2, and Table 1 (shown in different colors), denoted as $m_1, m_2, m_3$ (top-right), are retrieved from the database as shape priors to guide the selection of the NBV.
    After they are aligned with $\gamma$, a few viewpoints are sampled (middle-right). For each viewpoint, we compute its conditional information gain for $\gamma$.
    For the viewpoint marked in red circle, we compute the information gain of all visible voxels (middle-left). The prior entropy $H(x)$ and conditional entropies $H(x|m_i)$ according to the 3 model candidates, using 3 voxels as examples, are shown in the histograms (bottom) respectively.
    }
    \label{fig:NBV}
    \vspace{-7pt}
\end{figure}

\subsection{The next best view (NBV)}

Then, $\Omega$ moves to visit $\gamma$  and starts actively scanning  it with a plan of NBVs.
%
%

\paragraph{Idea}
%
Our idea is to take advantage of the candidates in $M(\gamma) = \{m_1, \cdots, m_{n_s} \} $ as shape priors to guide a full and complete scan of the object that $\gamma$ belongs to.
%
%
Thus, we expect to select the optimal view to achieve the maximal conditional information gain provided by $M(\gamma)$, which not only reduces the recognition uncertainty of $\gamma$ but also improves the discrimination rate of all the possible candidates.



\paragraph{Viewpoint candidates}
We uniformly sample $n_v$ points around the center $\gamma$, denoted as $\{ V_1, V_2, \cdots, V_{n_v}\}$, as the viewpoint candidates for NBVs (we set $n_v = 16$).
We aim to select the best viewpoint as the NBV with the largest conditional information gain. This provides better scans to increase the completeness  as well as the recognition rate of $\gamma$.
Note that there may be some invalid viewpoints which are invisible from $\gamma$ due to occlusions or the viewpoint is inside another object (e.g., wall). In these cases we just remove these viewpoints.

\paragraph{TDF representation of shapes}
%
We compute the information gain on the voxels around $\gamma$ with the assistance of the candidates in $M(\gamma)$. Thus, we first convert all of the shapes into the volumetric representations.
Specifically, for a shape $X \in \{\gamma\} \cup M(\gamma)$, we define the truncated distance function (TDF) on the voxels as:
\begin{equation}\label{equ:TDF}
f(x, X) =
\begin{cases}
 1, & x \in X, \\
 0, & \text{ otherwise, }
\end{cases}
\end{equation}
where $x$ denotes a voxel in the domain. To reduce the effect of data noise, we use a Gaussian blur to fuse the distance field to its neighboring voxels.
Moreover, we define a visibility function on the voxels from the viewpoint $V_j$ as: $g(x, V_j)=1$ if the voxel $x$ can be directly seen from $V_j$ without any occlusion; otherwise, $g(x, V_j)=0$.

\paragraph{Conditional information gain}
For each candidate viewpoint $V_j (j=1,\cdots,n_v)$ , we combine the conditional entropy and information gain as the conditional information gain for $\gamma$ with the help of all of the model candidates $m_i (i=1,\cdots,n_s)$ as (see Fig.~\ref{fig:NBV}):
\begin{equation}\label{equ:info-gain}
\max \limits_{j=1,\cdots,n_v} G^j = \sum_{i=1}^{n_s} p(m_i) G^j (m_i)
\end{equation}
where $p(m_i) = O(\gamma, m_i)/ \sum_{k=1}^{n_s} O(\gamma, m_k) $ is the probability of $m_i$ being the best candidate for $\gamma$ and
$G^j (m_i)$ is the information gain of all the visible voxels in $m_i$ from  $V_j$. We define $G^j (m_i)$ as:
\begin{equation}\label{equ:info-gain-pixels}
\sum \limits_{x \in \Delta} (H(x) - H(x | m_i))
\end{equation}
where $\Delta$ is the set of the voxels on $m_i$  but not on $\gamma$ and is visible from the viewpoint $V_j$, i.e.,
$$\Delta = \{x | f(x, m_i) \ne 0, f(x, \gamma)=0, g(x, V_j)=1 \},$$
\begin{equation}\label{equ:intial-uncertainty}
H(x) = - \sum_{k=1}^{n_s} p_x (m_k) \log p_x (m_k)
\end{equation}
is the prior entropy (initial uncertainty), \ca{and $p_x(m_k)$ is the prior probability defined on certain voxel $x \in X$ which is approximated by $p_x(m_k)=p(m_k)$.}
The uncertainty after observation is determined by conditional entropy as:
\begin{equation}\label{equ:certainty}
H(x | m_i) = p_x(0 | m_i) H_x (0) + p_x(1 | m_i) H_x (1)
\end{equation}
where
$p_x(1 | m_i) = f (x, m_i)$, $p_x(0 | m_i) = 1- p_x(1 | m_i)$, and
\begin{equation}\label{equ:certainty-2}
H_x (\delta) = \sum \limits_{k=1}^{n_s} {-p_x(m_k | \delta) \log p_x(m_k | \delta)}, \delta = 0, 1,
\end{equation}
where
\begin{equation}\label{equ:certainty-3}
p_x( m_i | \delta) = \frac{p_x (m_i) p_x (\delta | m_i)} {\sum \limits_{k=1}^{n_s} p_x (m_k) p_x (\delta | m_k)}, \delta = 0, 1.
\end{equation}

The quality of the viewpoint $V_j$ is defined as the above conditional information gain, which measures both the completeness and the recognition rate of the scan obtained from  $V_j$.
The viewpoint with the highest information gain is selected as the NBV.
\ca{This strategy tends to find as small number of views as possible for fine scanning and recognition of $\gamma$.}

\section{Results and Evaluation}
\label{sec:result}

We conducted a series of experiments, both with simulation in virtual scenes and with
robots running in real scenes, to evaluate the effectiveness of our method.
The evaluations are designed mainly around two questions regarding object-guided autoscanning:
1) How well our method is able to recognize the objects in the scene? and
2) How completely does our scan cover the objects?

\subsection{System and implementation}

\paragraph{Robotic system}
Our system runs with a customized robot platform with a single 6-DOF articulated arm holding a Microsoft Kinect RGB-D sensor, which is powered by the carried-on battery of the robot platform.
The robot has a built-in computer running a ROS system, which provides a package to enable standard robot behavior, such as navigation and arm actions.
Given a target view, the position and pose of the robot are computed, and an optimal collision-free smooth path is automatically planned by the package.
The system runs on a laptop PC with an Intel I7-6820HK CPU (quad core, 2.7GHz), 32GB RAM, and an Nvidia GeForce GTX 1080 graphics card carried by the robot and thus is self-contained and cableless, making it flexible enough for free movement (see the accompanying video).

%
%
%

\paragraph{Dataset}
The virtual scene dataset is built upon the scene datasets from SUNCG~\cite{song2016} and ScanNet~\cite{dai2017},
encompassing both human-modeled synthetic scenes (\ca{66} from SUNCG)
and human-scanned real scenes (\ca{38} from ScanNet).
The collection contains \ca{104} scenes spanning 5 categories, including bedrooms (\ca{21}),
sitting rooms (\ca{24}), kitchens (\ca{20}), etc.
Both datasets provide ground-truth object segmentation and labeling for the scenes.

\paragraph{Benchmark}
To facilitate quantitative evaluation of object-aware scene scanning,
we propose a benchmark, named the \emph{Object-Aware Scanning Benchmark (OASC)}, based on the virtual scene dataset.
\ca{Note, however, the scenes are mostly not composed with the objects from the 3D model database.}
We evaluate the performance of object-aware scanning from several aspects, including
object recognition, single-view object detection, object-level segmentation,
object coverage rate, and object coverage quality.
The corresponding metrics will be elaborated below, along with the quantitative evaluations and comparisons.
\ca{The benchmark dataset and the accompanying toolbox will be made publicly available.}


\paragraph{Objectness thresholds}
In our implementation, when a post-segmented object reaches an objectness score larger than 0.96, it is regarded as a full object, and thus is replaced by the most similar model from the database.
If  a post-segmented object has a very small objectness score, less than 0.05, we can regard it as noise in the scene and filter it out from $\mathcal{T}$.


\paragraph{Selection of similar model sets}
For the pre-segmented component $c\in \mathcal{C}$, we select the models most similar to it from the whole database $\mathcal{M}$ to construct $M(c)$ and use it in the post-segmentation to obtain post-segmented objects.
For the post-segmented object $r \in \mathcal{R}$, we select the models most similar to it from the subset of $\mathcal{M}$ that consists of only full models. This is because they provide  complete information for the original 3D models in $\mathcal{M}^*$.

\paragraph{Floor and wall}
The input scenes are regarded as flat layouts of floors, walls, and various furniture.
The floors and walls can be easily identified using some heuristic planar fitting methods. We exclude them from $\mathcal{T}$ for object identification and recognition.
We use the front boundaries of floors to guide $\Omega$ to move to the unexplored regions in case there are no visible objects within the viewable distance.
The scanning process terminates when there are no more objects in $\mathcal{T}$ and there are no more front boundaries on the floor.

\subsection{Performance and evaluation}

\paragraph{Performance of object recognition.}
To provide a quantitative evaluation of object recognition, we tested our method by simulating robotic
scanning in virtual scenes.
Both datasets provide ground-truth object segmentation and labeling for the scenes.
All the object categories that appear in the scenes are covered by our object database.
For each dataset, we randomly selected \ca{$7\sim14$} scenes for each \ca{scene} category.
Overly crowded scenes were avoided since they make obstacle-avoiding
robot navigation difficult, which is not the goal of this paper.
A sample of these scenes is shown in Fig.~\ref{fig:visrslt}.

In Table~\ref{tab:scenerecall}, we report the average recall and precision rate of object recognition
for each scene category. For each scene, the recall and precision are measured against
ground-truth objects at termination.
For all categories, our method obtains an average recall rate of \ca{65.93\%} and a precision rate of \ca{70.61\%}.
The recall rate for kitchens is relatively low, since most of the objects in these scenes
are small and placed on top of other objects, which makes them hard for our method to recognize.
In many living rooms, large furniture is mainly composed of planar surfaces (such as a long cabinet),
which are hard to characterize, leading to degraded precision rates in these scenes.
Otherwise, our method achieves  satisfactory recognition accuracy for objects placed on the floor.
The last two columns in the table show the recall and precision rates for ground objects only,
which are much higher than for all objects.
\ca{We also report the average recall and precision rate per object category in Table~\ref{tab:objrecall}.
Objects which are very sparse and difficult to recognize, such as towel rack and water dispenser,
are all put into the `Other' category.}

\begin{figure}[t!] \centering
	\begin{overpic}[width=1.0\linewidth,tics=10]{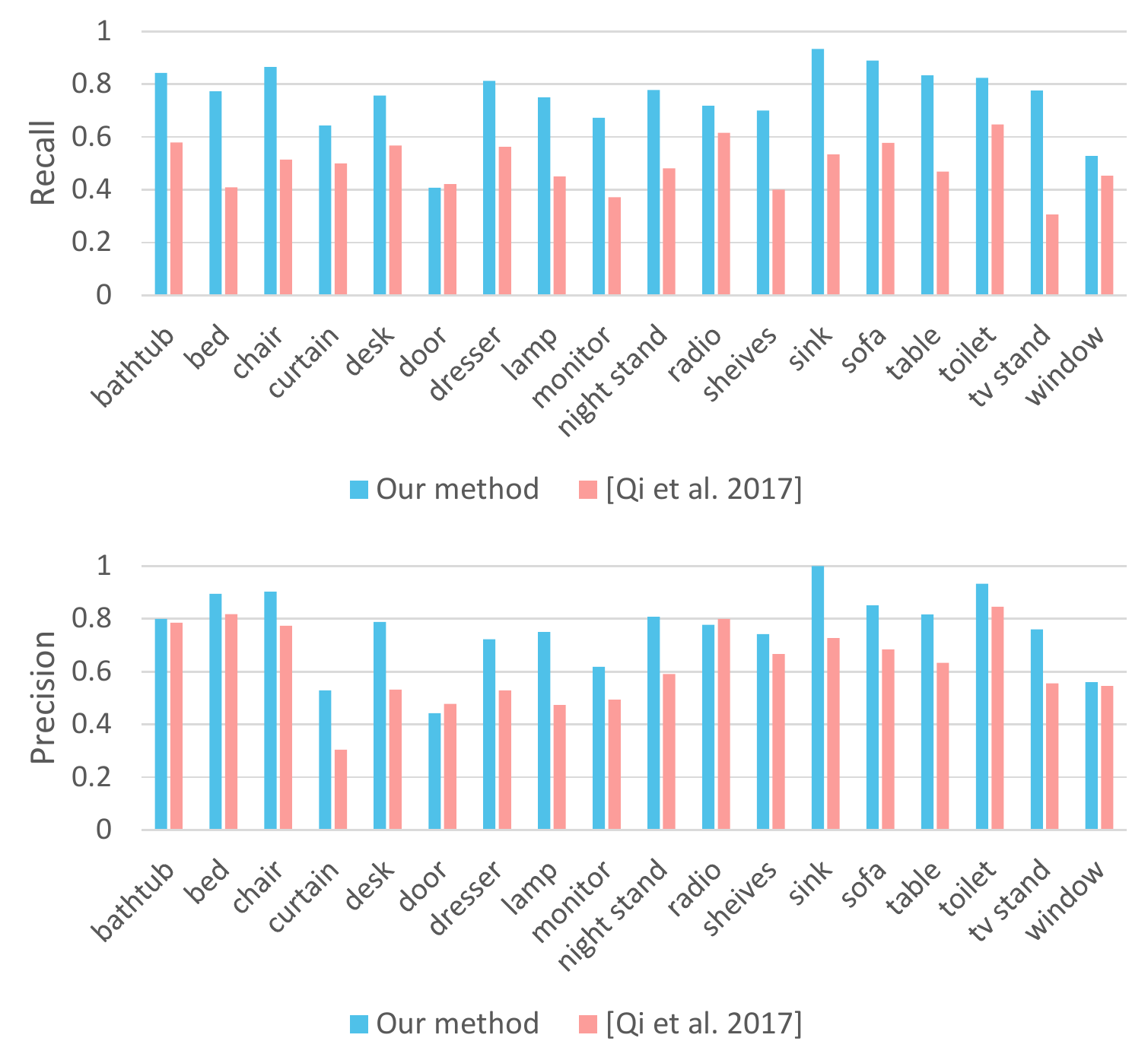}
	\end{overpic}
    \caption{\ca{Comparing object recognition with PointNet++~\cite{qi2017pointnetplusplus}.}}
    \label{fig:segrecog}
    \vspace{-7pt}
\end{figure}

\ca{We also compared our method with a state-of-the-art method, PointNet++~\cite{qi2017pointnetplusplus} for point cloud segmentation and labeling.
PointNet++ offers both object classification and scene parsing.
We first perform segmentation over the completely scanned scenes using the scene parsing function of PointNet++ and then classify the segments using the object classification function.
The comparison of average recall and precision for each object category can be found in Fig.~\ref{fig:segrecog}.
Our method provides a better performance due to the coupled solution to both segmentation and recognition.}

\begin{table}[!ht]
\caption{Performance of object recognition in each scene category. For each category, we report the number of scenes (\#S), the average number of objects per scene (\#O), and the recall and precision rates of object recognition.
The last two columns give the rates only for objects placed on the ground.}
\centering
\small
  \begin{tabular}{lcccccc}
    \hline
      Category & \#S & \#O & Recall & Precision & Recall(G) & Precision(G) \\
    \hline
    Bedroom         & 21 & 11 & 64.22\% & 86.13\% & 76.41\% & 93.13\%\\
    Living room     & 24 & 15 & 65.27\% & 58.40\% & 71.91\% & 60.21\%\\
    Kitchen         & 20 & 15 & 55.78\% & 66.65\% & 73.47\% & 76.82\%\\
    Bathroom        & 19 &  6 & 81.74\% & 84.68\% & 83.19\% & 85.45\%\\
    Office          & 20 & 11 & 74.77\% & 78.81\% & 85.56\% & 87.91\%\\
    \hline
  \end{tabular}
\label{tab:scenerecall}
\end{table}

\begin{table}[!ht]\ca{
\caption{\ca{Performance of object recognition in each object category. For each category, we report the number of objects (\#O), and the recall and precision rates of object recognition.}}
\centering
\scalebox{0.8}{
  \begin{tabular}{lccc|lccc}
    \hline
    Category & \#O & Recall & Precision & Category & \#O & Recall & Precision \\
    \hline
    Bathtub     &  19 & 84.21\% & 80.0\%  & Night stand  &  27 & 77.78\% & 80.77\%\\
    Bed         &  22 & 77.27\% & 89.47\% & Radio        &  39 & 71.79\% & 77.78\%\\
    Cabinet     &  84 & 63.10\% & 53.54\% & Refrigerator &  12 & 58.33\% & 63.64\%\\
    Can         &   9 & 100.0\% & 75.0\%  & Sheives      &  70 & 70.0\%  & 74.24\%\\
    Chair       & 140 & 86.43\% & 90.30\% & Sink         &  15 & 93.33\% & 100.0\%\\
    Curtain     &  14 & 64.29\% & 52.94\% & Sofa         &  45 & 88.89\% & 85.11\%\\
    Desk        &  74 & 75.68\% & 78.87\% & Table        &  96 & 83.33\% & 81.63\%\\
    Door        &  76 & 40.79\% & 44.28\% & Toilet       &  17 & 82.35\% & 93.33\%\\
    Dresser     &  16 & 81.25\% & 72.22\% & TV stand     &  49 & 77.55\% & 76.0\%\\
    Lamp        &  20 & 75.0\%  & 75.0\%  & Window       &  53 & 52.83\% & 56.0\%\\
    Moniter     & 113 & 67.26\% & 61.79\% & Other        & 211 & 33.18\% & 56.45\%\\
    \hline
  \end{tabular}
}
\label{tab:objrecall}}
\end{table}

\paragraph{Performance on single-view object detection.}
A key factor of object-aware guidance is the ability to detect potential
objects and assess their significance based on  actively acquired partial observations.
To evaluate this ability, we tested our method for \emph{single-view} object detection and evaluated
the performance based on the image-based objectness metric proposed in~\cite{alexe2012}.
It measures the objectness of a detection window in 2D images, based on the Intersection of Union (IoU)
against the ground-truths object bounding box.
Since our method performs detections in 3D, the objectness measure is computed by projecting
the 3D detection back to the input views of images.
The experiment is conducted on the 2D-3D-Semantics dataset~\cite{armeni2017}, which provides human annotated ground-truth for object detection.

\begin{figure}[t!] \centering
	\begin{overpic}[width=0.8\linewidth,tics=10]{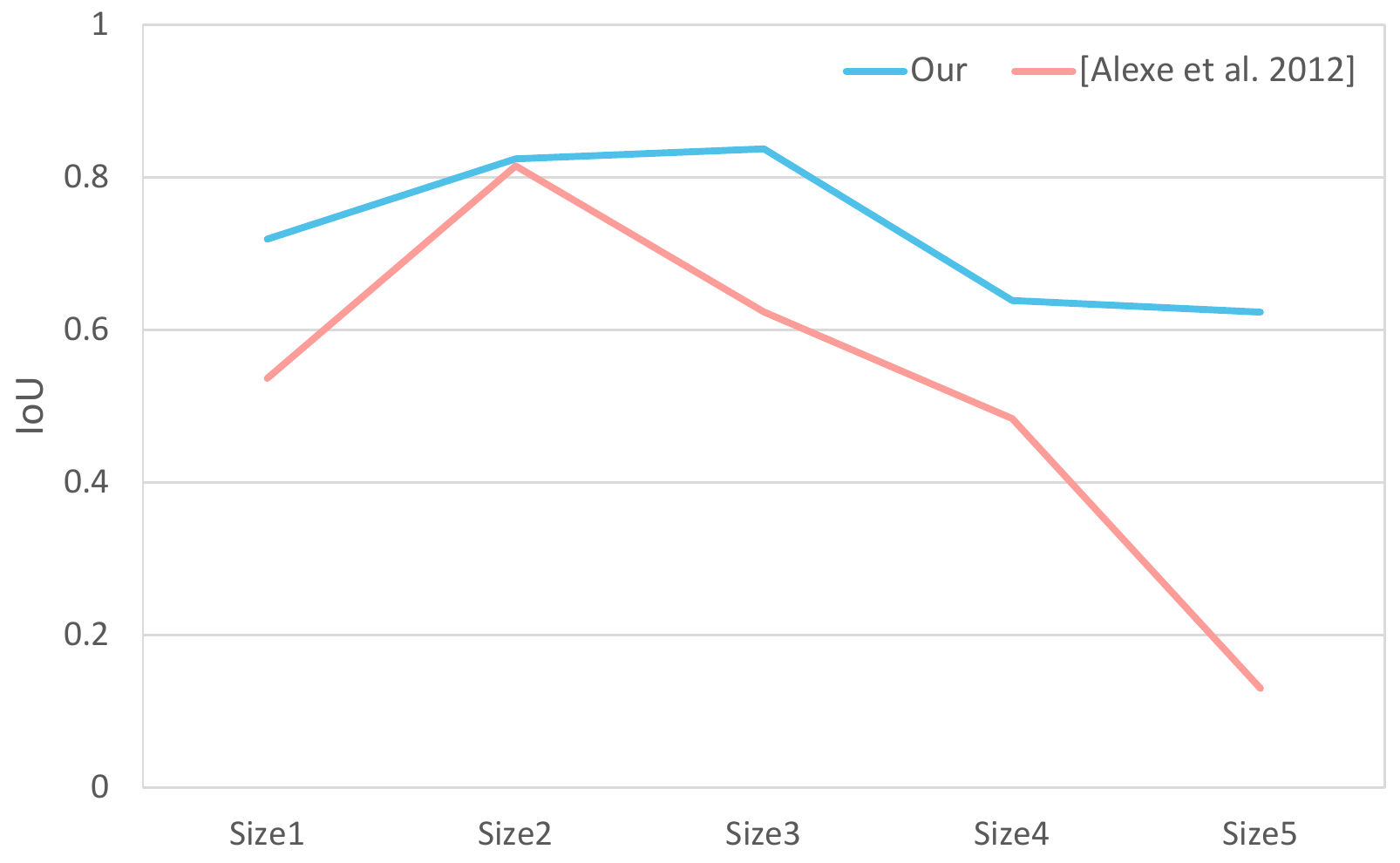}
	\end{overpic}
    \caption{Plots of objectness measurements over increasing object sizes (measured by the volume of bounding box).}
    \label{fig:objectness}
    \vspace{-7pt}
\end{figure}

Fig.~\ref{fig:objectness} plots the objectness measure over the size of the objects (measured
by the volume of bounding box).
We also plot the curves for the image-based object detection method in~\cite{alexe2012}.
Note that the image-based method is plotted only as a reference rather than for the purpose of comparison,
since our method also utilizes depth information.
In general, objects that are too small or too large are difficult to detect, as we have mentioned in the previous evaluation.
Nevertheless, our method performs well on a wider range of object sizes.

\paragraph{Performance of object-level segmentation.}
To test the ability of our method on object segmentation,
we evaluate the object-wise segmentation quality on the SunCG and ScanNet datasets.
Both datasets have ground-truth object segmentation.
For each detected object, we measure the Rand Index~\cite{Chen2009}
of the segmentation against its ground-truth.
\ca{The Rand Index for point cloud segmentation is defined as:}
\ca{
$$
RI(S_1,S_2) = {\begin{pmatrix} 2 \\ n \end{pmatrix}}^{-1} \sum_{i,j,i<j}[C_{ij} P_{ij} + (1 - C_{ij})(1 - P_{ij})],
$$
}\ca{where $S_1$ and $S_2$ are two segmentations, and $n$ is the number of points in the point cloud being segmented.
Suppose $s_i^1$ and $s_i^2$ are the segment IDs of point $i$ in $S_1$ and $S_2$,
then $C_{ij} = 1$ iff. $s_i^1 = s_j^1$, and $P_{ij} = 1$ iff. $s_i^2 = s_j^2$.}
Since our method performs virtual scanning and surface reconstruction for the
test scenes, the Rand Index is estimated by first aligning the reconstructed scene against the
ground-truth scene and then transfering the ground-truth segmentation onto the reconstructed surface.
Our method is compared to~\cite{tateno2015}, which is an incremental object segmentation approach based on dense SLAM, \ca{and to~\cite{qi2017pointnetplusplus}, a learning-based semantic segmentation method.}

\begin{figure}[t!] \centering
	\begin{overpic}[width=1.0\linewidth,tics=10]{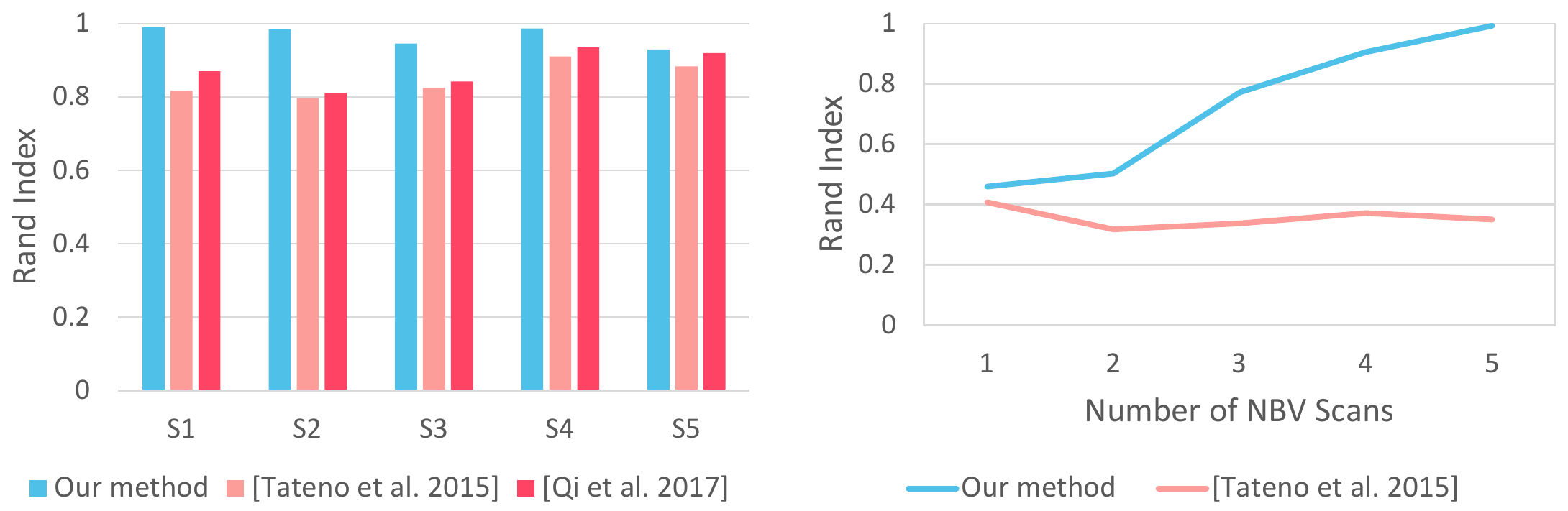}
    \put(20,-3){\small (a)}
    \put(70,-3){\small (b)}
	\end{overpic}
    \caption{(a): Comparing Rand Index of segmentation between our method, \pcite{tateno2015} \ca{and \pcite{qi2017pointnetplusplus}}, over different scene categories. (b): Rand Index over increasing number of NBV scans.}
    \label{fig:segplot}
    \vspace{-7pt}
\end{figure}

Fig.~\ref{fig:segplot}(a) shows the Rand Index for the \ca{three} methods, over the five
scene categories listed in Table~\ref{tab:scenerecall}. \ca{Hereafter, we refer to the five categories as S1$\sim$S5 for short.} Our method consistently offers better segmentation quality in all scenes.
In Fig.~\ref{fig:segplot}(b), we evaluate the segmentation performance over an increasing number of
NBV scans for \ca{our method and \cite{tateno2015}}.
The plot shows that the segmentation accuracy of the two methods is similar at the beginning, but our method
improves much faster as more NBV scans come in. This is because once an object is recognized, the segmentation
is improved significantly, thus verifying the effect of our object-based segmentation.
In addition, our method achieves a high Rand Index ($>0.8$) using an average of four NBV scans, which
also demonstrates the effectiveness of our active object recognition and segmentation.

\paragraph{Object coverage rate and quality.}
Object-aware scanning is naturally concerned with the scan coverage of objects.
We measure the coverage rate and quality of scene objects during or after autonomous scanning.
The \emph{object-wise coverage rate} measures the per-object \emph{valid coverage} by measuring the surface area
of an object, which is visible to robotic views within a valid scanning range.
In practice, this can be done by counting the surface voxels within the valid scanning range in the volumetric representation (we use OctoMap~\cite{hornung2013} in our implementation) of the virtual scenes.
The coverage rate can then be computed as:
$$
R_\text{cover} = \frac{1}{|\mathcal{V}_\text{S}|} \int_{v \in \mathcal{V}_\text{S}}{\delta_\text{detect}(v)\cdot \delta_\text{vis}(v)},
$$
where $\mathcal{V}_\text{S}$ is the set of surface voxels for all the ground-truth objects.
$\delta_\text{detect}$ and $\delta_\text{vis}$ are Dirac delta functions that indicate whether a voxel $v$
is on a detected object and if it is visible to the scanner within a valid scanning range.
\emph{Object-wise coverage quality} is measured similarly, except that
a quality measure is computed for each voxel and the measures for all voxels are normalized as:
$$
Q_\text{cover} = \frac{1}{|\mathcal{V}_\text{S}|} \int_{v \in \mathcal{V}_\text{S}}{\delta_\text{detect}(v)\cdot \delta_\text{vis}(v) \cdot q(v)},
$$
where $q(v)$ measures the scanning quality towards a voxel $v$, accounting for both the scanning
distance and the angle:
$$
q(v) = e^{-\frac{\theta^2(\textbf{n}_v,\textbf{d}_{cv})}{\sigma^2 \theta_\text{t}}} \cdot e^{-\frac{(d(v,c)-d_\text{min})^2}{\sigma^2 d_\text{t}}},
$$
where $\theta(\textbf{n}_v,\textbf{d}_{cv})$ is the angle between the normal at  surface
voxel $v$ and the viewing vector from camera $c$ to $v$.
$d(v,c)$ is the distance between camera $c$ and voxel $v$.
$d_\text{min}$ is the minimum distance for valid scanning.
We set $\sigma=0.6$, $\theta_\text{t}=\pi/2$, and $d_\text{t}=d_\text{max}$, with
$d_\text{max}$ being the maximum distance for valid scanning.

\begin{figure}[t!] \centering
	\begin{overpic}[width=1.0\linewidth,tics=10]{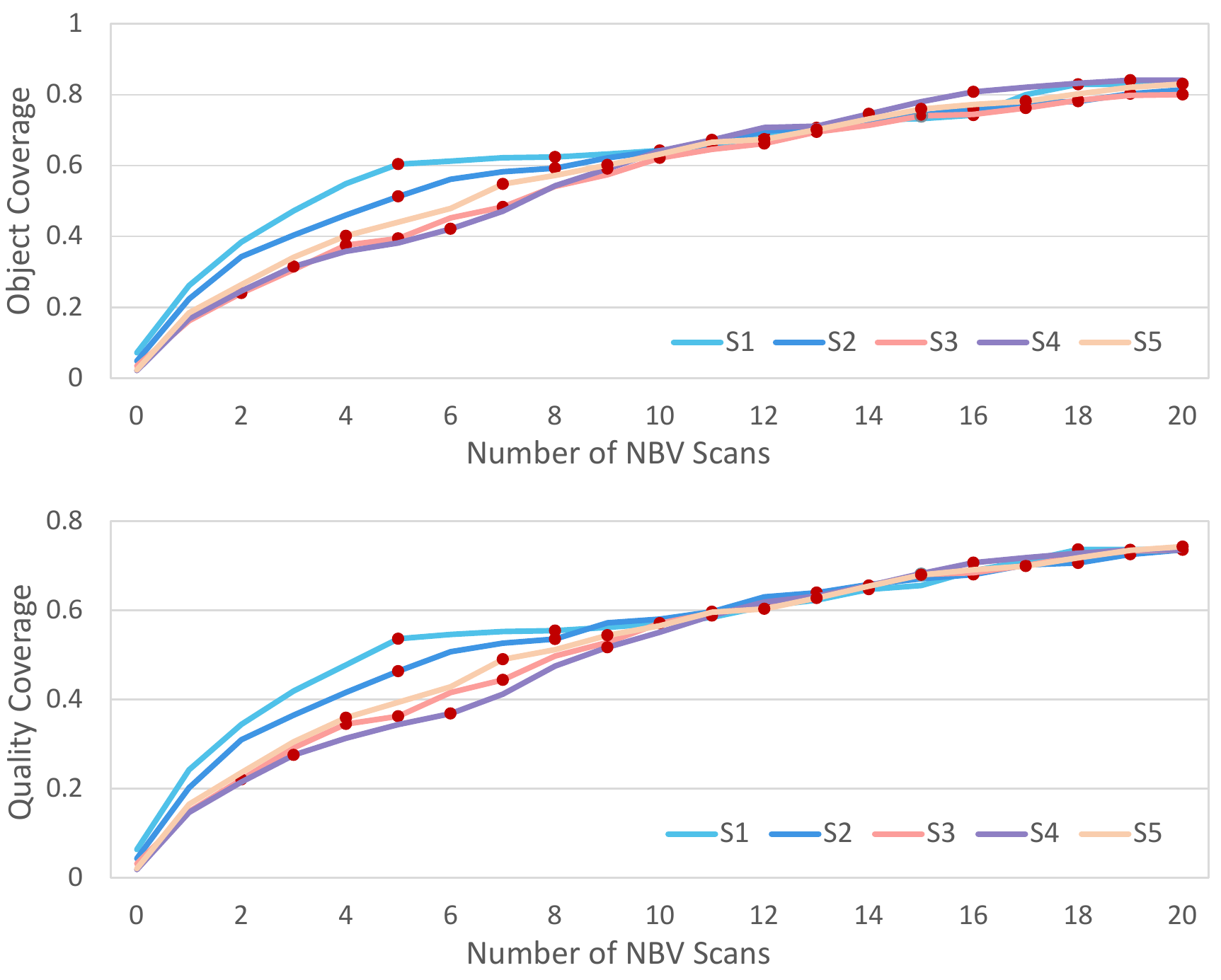}
	\end{overpic}
    \caption{Coverage rate and quality over increasing number of NBOs and NBVs, over five different scene types.
Along the curves, NBO points are marked with red dots.}
    \label{fig:coverage-nbv}
    \vspace{-7pt}
\end{figure}

Fig.~\ref{fig:coverage-nbv} shows the coverage rate and quality over an increasing number of NBOs and NBVs, for five different scene types.
Along the curves, the points at which the scanning is switched to the next object (NBO) are marked with red dots.
Our method results in satisfactory coverage ($>80\%$) with $1\sim5$ NBVs per object.
An interesting phenomenon to notice is that the required number of per-object NBVs decreases as
the scanning proceeds, since some objects may be scanned incidentally while others are being visited and scanned.
This feature of our method greatly improves the scanning efficiency:
the more objects there are in a scene, the more scanning effort can be saved.
Furthermore, our method achieves high quality coverage due to the consideration of scanning distance
and view angle. In the ablation studies below, we test the effect of these algorithmic settings
in more detail.

\begin{figure}[t!] \centering
	\begin{overpic}[width=1.0\linewidth,tics=10]{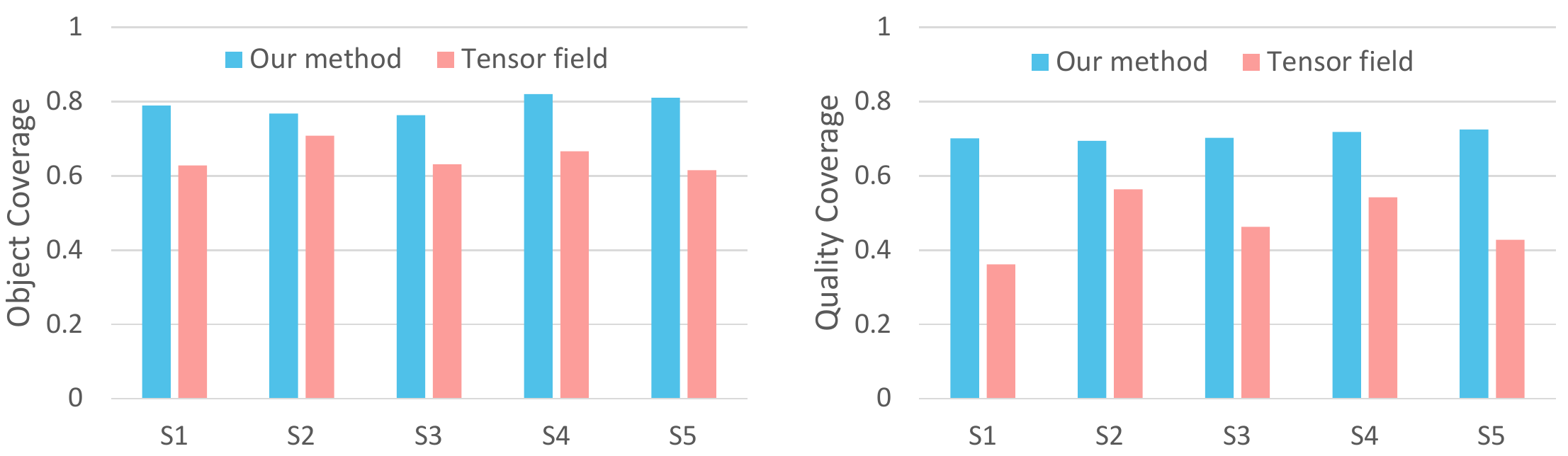}
	\end{overpic}
    \caption{Comparing object coverage rate and quality against tensor field guided autoscanning~\cite{Xu2017}.}
    \label{fig:coverage}
    \vspace{-7pt}
\end{figure}

Fig.~\ref{fig:coverage} compares the coverage rate and quality over different scene categories
with a baseline method, i.e., tensor field guided autonomous scene reconstruction~\cite{Xu2017}.
This method is not designed for object-aware scanning, as it leads to lower values in our object-aware metrics for all of the tested scenes.

\begin{figure}[t!] \centering
	\begin{overpic}[width=1.0\linewidth,tics=10]{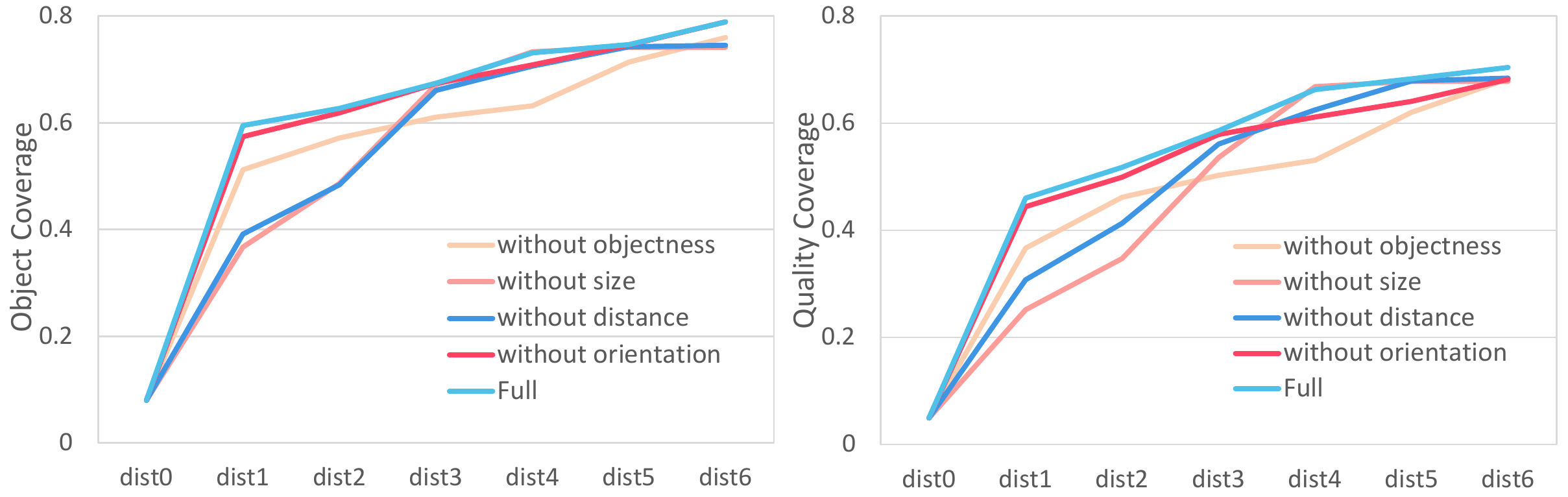}
	\end{overpic}
    \caption{Ablation studies on different terms in NBO estimation.}
    \label{fig:ablation}
    \vspace{-7pt}
\end{figure}

\paragraph{Ablation studies on NBO estimation.}
To justify the specific design choices in our NBO estimation,
we compare our full method against a series of ablated baselines that remove various energy terms including
the objectness term, the size term, the distance term, and the orientation term.
For each method,
we evaluate the efficiency of the estimated NBOs by measuring the object-wise coverage rate
over the travel distance of the robot, as well as object-wise coverage quality over the number of NBOs.
Fig.~\ref{fig:ablation} plots the results for five scene categories, demonstrating that
our method achieves the most efficient object-aware scanning.
From the results, it can be observed that the most critical terms are objectness and size.
This directly verifies the usefulness of objectness-driven scanning.
Besides, the size term, which is used to prioritize more prominent objects of larger size, also heavily contributes to the overall scanning efficiency.
%

\begin{figure}[t!] \centering
	\begin{overpic}[width=0.9\linewidth,tics=10]{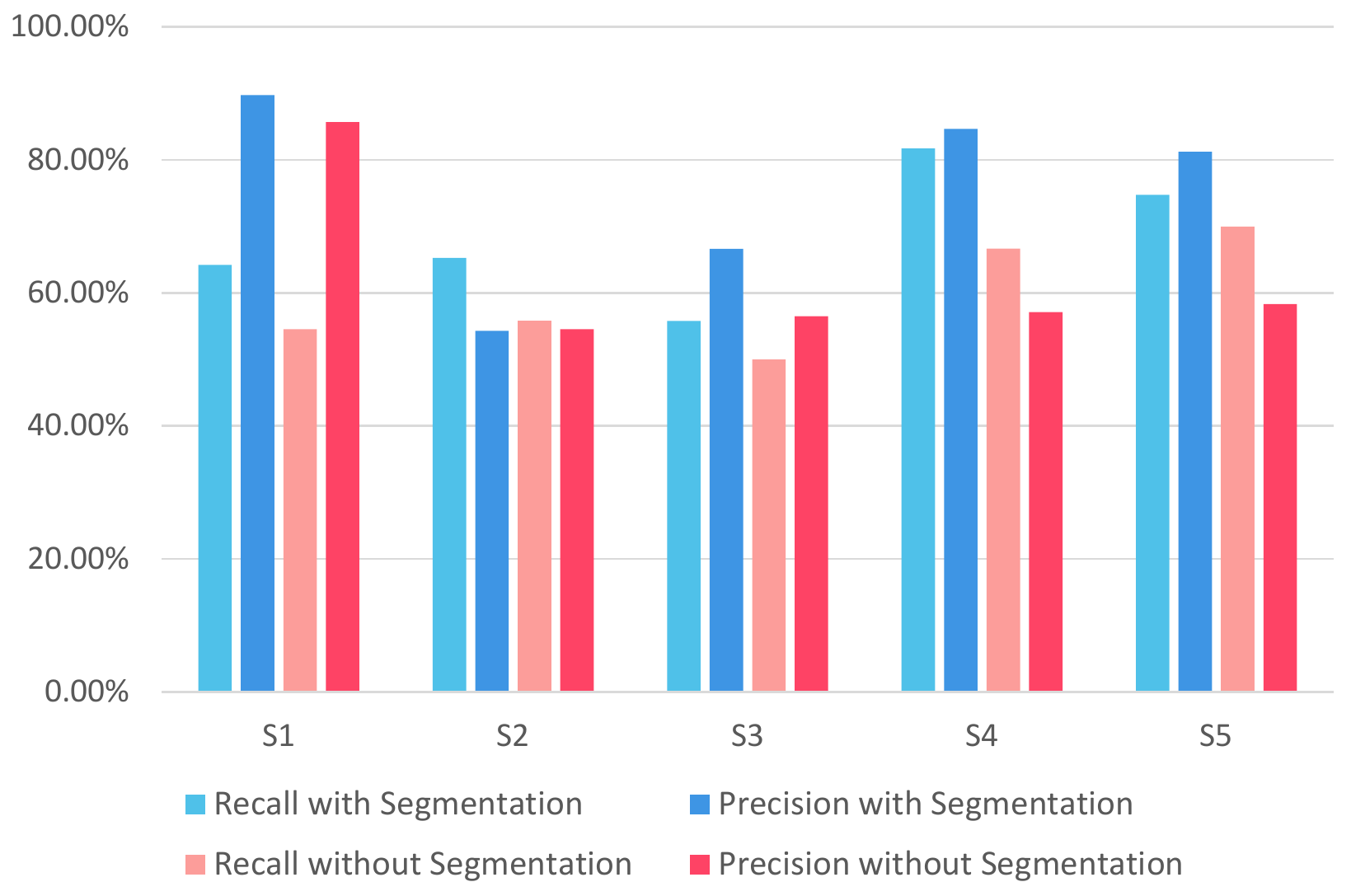}
	\end{overpic}
    \caption{Effect of database pre-segmentation on object recognition performance.}
    \label{fig:datasetseg}
    \vspace{-7pt}
\end{figure}

\paragraph{The effect of pre-segmentation of database shapes.}
Our database shapes are pre-segmented using exactly the same method
we adopt in online segmentation. This promotes more efficient online object detection and object segmentation.
In Fig.~\ref{fig:datasetseg}, we evaluate the effects of  pre-segmentation on
object recognition, over different scene categories.
From the plot, it can be seen that our method, which includes pre-segmentation, performs better than methods lacking this important step.

\subsection{Experiments}

\begin{figure*}[t!] \centering
	\begin{overpic}[width=1.0\linewidth,tics=10]{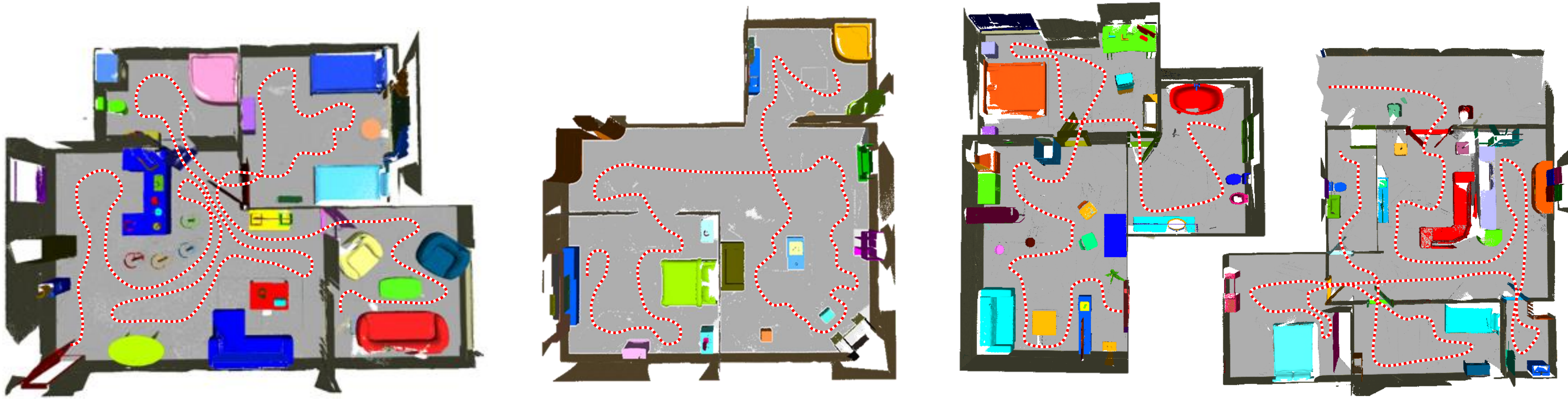}
	\end{overpic}
    \caption{Visual results of object-aware scanning for virtual simulation.}
    \label{fig:visrslt}
    \vspace{-7pt}
\end{figure*}

\paragraph{Virtual simulation}
Fig.~\ref{fig:visrslt} shows the visual results of our object-guided scanning for virtual simulation.
For virtual simulation, we implement our method in the robotic simulation environment of
Gazebo~\shortcite{Gazebo} running on top of ROS. The virtual scenes are from both SUNCG and ScanNet.
For each example, we show the results of object recognition by visualizing
the top view of the recognized objects, indicated with bright colors, and the unrecognized content, shown in grey.
The robot's path for navigation and scanning is also shown.
From the visualization, it can be observed that our method achieves a one pass
scan of the scenes that is guided by object recognition, along simple paths of motion.

\begin{figure*}[t!] \centering
	\begin{overpic}[width=1.0\linewidth,tics=10]{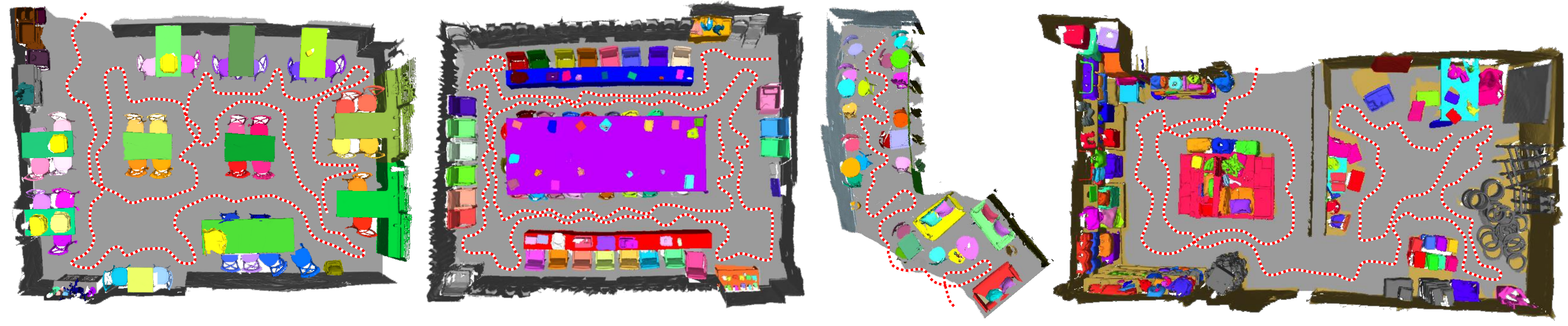}
	\end{overpic}
    \caption{Visual results of object-aware scanning for real running.}
    \label{fig:realrslt}
    \vspace{-7pt}
\end{figure*}

\paragraph{Real robot tests}
We have tested our robot by scanning four unknown indoor rooms found around our department,
including one coffee house, one meeting room, one resting room, and one small shop.
Fig.~\ref{fig:realrslt} shows the visual results of object-aware reconstruction on these scenes.
Based on a manually counted ground-truth, our method successfully performs the object recognition task:
coffee house (recall: 61.90\%; precision: 72.22\%), meeting room (recall: 63.64\%; precision: 62.89\%),
resting room (recall: 57.69\%; precision: 65.22\%) and shop (recall: 48.15\%; precision: 54.17\%).

\paragraph{Timing}
Table~\ref{tab:time} reports the online running time of our method, as well as  for the various algorithmic components including navigation, segmentation, and the computation of NBOs and NBVs.
Besides navigation, the main speed bottleneck of our method is segmentation, where the graph cuts require
$O(n)$ per-component feature extraction and matching for data term estimation, and $O(n^2)$ for smoothness term
computation. Here, $n$ is the size of the current set of pre-segmented components.
The feature matching has linear complexity with respect to the size of the component database.

\begin{table}[!ht]
\caption{Running time (in minutes) of our method and its various algorithmic components, including navigation, segmentation, and the computation of NBOs and NBVs. Timings for virtual scenes (V) are averaged for scenes in each scene category. The scanning of the three real scenes (R) was timed during the robot's real running time.}
\centering
\small
  \begin{tabular}{lccccc}
    \hline
      Category & Total & Navigate & Segment & NBO & NBV\\
    \hline
    Bedroom (V)         & 47.8 & 24.1 & 20.1 & 2.0 & 1.6\\
    Living room (V)     & 57.0 & 30.4 & 22.2 & 2.3 & 2.1\\
    Kitchen (V)         & 37.5 & 16.2 & 17.6 & 2.0 & 1.7 \\
    Bathroom (V)        & 29.5 & 14.8 & 12.2 & 1.3 & 1.2 \\
    Office (V)          & 40.8 & 21.3 & 16.0 & 1.9 & 1.6\\
    Meeting room (R)    & 101.4 & 62.3 & 32.4 & 3.6 & 3.1\\
    Resting room (R)    & 78.5 & 47.9 & 25.4 & 2.9 & 2.3\\
    Office (R)          & 94.7 & 56.9 & 30.3 & 4.2 & 3.3\\
    \hline
  \end{tabular}
\label{tab:time}
\end{table}

\section{conclusions}
\label{sec:future}

We present an object-guided approach for autonomous scene exploration, reconstruction, and understanding.
At the core of our approach is a series of object driven algorithms.
First, a model-driven objectness is defined and used to measure the similarity and completeness of pre-segmented components.
Second, an objectness-based segmentation is developed to obtain a set of post-segmented objects.
Then, we adopt an objectness-based NBO strategy to identify and recognize the OOI with the largest objectness score and visual saliency  to allow the robot to start actively scanning the OOI.
A small number of views are computed by the NBV method to guide the scanning. This guarantees that the OOI is scanned fully and completely.
%
After the robot identifies the OOI, one could replace it with the 3D model
which is the most similar to it, retrieved from the database. Then it moves to the next OOI and starts to scan it.
This repeats until all of the objects in the scene have been scanned and reconstructed, resulting in a full reconstruction of the scene.
We have shown a large number of experimental results and comparison results to validate our proposed approach and prove its feasibility and effectiveness.

\paragraph{Limitations}
Our current solution for autoscanning suffers from several limitations.
%
%
First, our approach does not work well for highly cluttered scenes with small shapes.
The quality of acquired data and the segmentations decreases significantly due to mis-segmentation and occlusion when there is severe object clutter in the scene.
Second, the precision and stability of recognizing and identifying 3D objects highly depend on the database that the 3D models come from as the database provides the prior knowledge necessary for the robots to understand 3D shapes.
If there are no similar models in the database, then the recognition rate may be very low, resulting in incorrect exploration and reconstruction. %
Third, our approach uses geometric information from single objects only in order to guide the exploration and reconstruction. It
does not infer  high-level semantics, such as spatial relationships among multiple objects or group structures of similar or functional objects.
These kinds of higher level guidance may produce preferable navigation paths, such as anticipation of accessibility or safety.

\paragraph{Future work}
Our work on object-aware autoscanning opens up an inspiring direction in both robotics and graphics communities. We believe it will inspire promising researches in the future.
%
First, we rely on existing methods for several technical components of our approach, such as pre-segmentation, 3D shape descriptors, and partial matching. These are fundamental tasks and play crucial roles in 3D data processing and scene analysis. These tasks deserve more in-depth and theoretical study.
Second, it is worth studying learning-based methods for scene exploration and reconstruction via state-of-the-art deep learning techniques. Construction of the training set and the deep network are both challenging.
Third, it is interesting to combine our method with image-based methods because images provide fine grain information, such as color, texture, and lighting, which may aid recognition of the objects in the scene.
\ca{Lastly, extending our approach to explore and reconstruct outdoor scenes with LiDAR sensors or drones is a prospective research direction. The popularity of drones and self-driving cars are stimulating more advanced research and applications for automatic exploration of unknown scenarios and scenes.
}

\section*{Acknowledgements}
We thank the anonymous reviewers for their valuable comments. This work was supported in part by
NSFC (61672482, 61672481, 61622212, 61572507, 61532003, 61522213),
the One Hundred Talent Project of the Chinese Academy of Sciences,
973 Program (2015CB35\\2501),
Guangdong Science and Technology Program (2015A03031201\\5) and
Shenzhen Science and Technology Program (KQJSCX20170727\\101233642, JCYJ20151015151249564).

\bibliographystyle{ACM-Reference-Format}
\bibliography{nbo}

\end{document}